\documentclass[resetfootnote,trackchanges,twocolumn,twocolappendix]{aastex701}

\usepackage{tabularx}
\usepackage{amsmath}
\usepackage{tablefootnote}
\usepackage{placeins}
\usepackage{enumitem}      
\usepackage{array}         
\usepackage{booktabs}        
\usepackage{xcolor}
\usepackage{colortbl}
\usepackage{multirow}
\usepackage{longtable}
\usepackage{booktabs}
\newcommand{\lya}{Ly$\alpha$}

\newcommand{\HI}{\rm H\,{\textsc {i}}}

\newcommand{\SiII}{\rm Si\,{\textsc {ii}}}
\newcommand{\SiIII}{\rm Si\,{\textsc {iii}}}
\newcommand{\SiIV}{\rm Si\,{\textsc {iv}}}
\newcommand{\CII}{\rm C\,{\textsc {ii}}}
\newcommand{\CIV}{\rm C\,{\textsc {iv}}}
\newcommand{\MgII}{\rm Mg\,{\textsc {ii}}}

\definecolor{gold}{RGB}{255,166,0}   
\definecolor{silver}{RGB}{138,138,138}
\definecolor{bronze}{RGB}{150,75,0}

\begin{document}

\title{The Prevalence of Turbulence-Regulated Multiphase Galactic Winds in Star-Forming Galaxies}

\author[0000-0001-5113-7558]{Zhihui Li}
\affiliation{Center for Astrophysical Sciences, Department of Physics \& Astronomy, Johns Hopkins University, Baltimore, MD 21218, USA}
\email[show]{zli367@jh.edu}

\author[0000-0001-6670-6370]{Timothy Heckman}
\affiliation{Center for Astrophysical Sciences, Department of Physics \& Astronomy, Johns Hopkins University, Baltimore, MD 21218, USA}
\email{theckma1@jhu.edu}

\author[0000-0003-2491-060X]{Max Gronke}
\affiliation{Max-Planck Institute for Astrophysics, Karl-Schwarzschild-Str. 1, D-85741 Garching, Germany}
\affiliation{Astronomisches Rechen-Institut, Zentrum f\"{u}r Astronomie, Universit\"{a}t Heidelberg, Mönchhofstra$\beta$e 12-14, 69120 Heidelberg, Germany}
\email{max.gronke@uni-heidelberg.de}

\author[0000-0002-9217-7051]{Xinfeng Xu}
\affiliation{Department of Physics and Astronomy, Northwestern University, 2145 Sheridan Road, Evanston, IL, 60208, USA}
\affiliation{Center for Interdisciplinary Exploration and Research in Astrophysics (CIERA), Northwestern University, 1800 Sherman Avenue, Evanston, IL, 60201, USA}
\email{xinfeng.xu@northwestern.edu}

\author[0000-0002-6586-4446]{Alaina Henry}
\affiliation{Center for Astrophysical Sciences, Department of Physics \& Astronomy, Johns Hopkins University, Baltimore, MD 21218, USA}
\affiliation{Space Telescope Science Institute, 3700 San Martin Drive, Baltimore, MD 21218, USA}
\email{ahenry@stsci.edu}

\author[0000-0001-9735-7484]{Evan Schneider}
\affiliation{Department of Physics \& Astronomy and PITT-PACC, University of Pittsburgh, 100 Allen Hall, 3941 O’Hara Street, Pittsburgh, 15260, PA, USA.}
\email{eschneider@pitt.edu}

\author[0000-0002-7918-3086]{Matthew Abruzzo}
\affiliation{Department of Physics \& Astronomy and PITT-PACC, University of Pittsburgh, 100 Allen Hall, 3941 O’Hara Street, Pittsburgh, 15260, PA, USA.}
\email{mwa2113@columbia.edu}

\author[0000-0002-4153-053X]{Danielle Berg}
\affiliation{Department of Astronomy, The University of Texas at Austin, 2515 Speedway, Stop C1400, Austin, TX 78712, USA}
\email{daberg@austin.utexas.edu}

\author[0000-0003-4372-2006]{Bethan James}
\affiliation{Space Telescope Science Institute, 3700 San Martin Drive, Baltimore, MD 21218, USA}
\email{bjames@stsci.edu}

\author[0000-0001-9189-7818]{Crystal Martin}
\affiliation{Department of Physics, University of California, Santa Barbara, CA 93106, USA}
\email{cmartin@physics.ucsb.edu}

\author[0000-0002-0302-2577]{John Chisholm}
\affiliation{Department of Astronomy, The University of Texas at Austin, 2515 Speedway, Stop C1400, Austin, TX 78712, USA}
\email{chisholm@austin.utexas.edu}

%% Use the \collaboration command to identify collaborations. This command
%% takes an optional argument that is either a number or the word "all"
%% which tells the compiler how many of the authors above the command to
%% show. For example "\collaboration[all]{(DELVE Collaboration)}" wil include
%% all the authors above this command.
%%
%% Mark off the abstract in the ``abstract'' environment. 
\begin{abstract}

We build upon our previously developed multi-ion radiative transfer (RT) framework, \texttt{PEACOCK}, to investigate the kinematic and energetic structure of cool–to-warm galactic winds in a sample of 50 nearby star-forming galaxies. Using self-consistent constraints derived from joint modeling of \lya\ and multiple ultraviolet metal lines, we analyze how bulk outflows and turbulent motions contribute to the dynamics and energy budget of the galactic wind in the circumgalactic medium (CGM). We find that macroscopic turbulent velocities are frequently comparable to, and in some systems exceed, the coherent bulk outflow velocity. The associated turbulent pressure often dominates over both microscopic pressure and ram pressure, indicating that turbulence constitutes a primary contributor to the kinetic energy budget of the CGM wind. Wind kinematics, ionic column densities, and metal mass outflow rates all scale systematically with stellar mass and star formation rate, demonstrating a strong coupling between stellar feedback and CGM structure. Incorporating turbulent motions strengthens these CGM–galaxy scaling relations and favors an energy-driven feedback regime. The total kinetic energy flux of the cool–to-warm CGM correlates tightly with the mechanical energy injection rate from star formation, implying that stellar feedback provides sufficient power to sustain both coherent outflows and turbulence. Comparisons with phenomenological line profile fitting methods further show that simplified treatments can introduce systematic biases in inferred wind properties. Taken together, these results support a turbulence-regulated picture of galactic winds in which a substantial fraction of feedback energy is stored in turbulent motions within a multiphase CGM.

\end{abstract}

%% Keywords should appear after the \end{abstract} command. 
%% The AAS Journals now uses Unified Astronomy Thesaurus (UAT) concepts:
%% https://astrothesaurus.org
%% You will be asked to selected these concepts during the submission process
%% but this old "keyword" functionality is maintained in case authors want
%% to include these concepts in their preprints.
%%
%% You can use the \uat command to link your UAT concepts back its source.

%% From the front matter, we move on to the body of the paper.
%% Sections are demarcated by \section and \subsection, respectively.
%% Observe the use of the LaTeX \label
%% command after the \subsection to give a symbolic KEY to the
%% subsection for cross-referencing in a \ref command.
%% You can use LaTeX's \ref and \label commands to keep track of
%% cross-references to sections, equations, tables, and figures.
%% That way, if you change the order of any elements, LaTeX will
%% automatically renumber them.

\section{Introduction} 

Galactic-scale winds are a fundamental component of galaxy evolution, regulating star formation, redistributing metals, and mediating the exchange of mass and energy between galaxies and their surrounding halos \citep{Veilleux2005, Tumlinson2017, Thompson2024}. These winds are commonly observed through ultraviolet (UV) emission and absorption lines that trace multiphase gas spanning a wide range of densities, temperatures, and ionization states \citep[e.g.,][]{Heckman2000, Steidel2010, Rubin2014}. Over the past two decades, spectroscopic surveys have established that galactic outflows are ubiquitous across cosmic time and are intimately connected to stellar feedback processes \citep[e.g.,][]{Heckman2000, Shapley2003, Veilleux2005, Steidel2010, Rubin2014, Heckman2017, Tumlinson2017}.

Despite this substantial observational progress, the dynamical structure and energy partition of galactic winds in the circumgalactic medium (CGM) remain incompletely understood. Most empirical analyses infer characteristic outflow velocities and mass-loading factors under simplified geometric assumptions, often modeling winds as coherent, large-scale bulk flows \citep[e.g.,][]{Heckman2000, Steidel2010, Chisholm2015, Xu2022}. In this conventional framework, stellar feedback is assumed to channel energy primarily into directed outflows that transport mass and momentum away from galaxies \citep[e.g.,][]{Murray05, Veilleux2005, Heckman2015}, while turbulent motions are typically treated as secondary by-products of hydrodynamic instabilities. As a result, the relative partition of feedback energy between coherent bulk motion and stochastic turbulent motions is rarely quantified in a self-consistent manner. Such treatments may therefore overlook the possibility that turbulence itself constitutes a dynamically significant reservoir of energy within the multiphase wind \citep[e.g.,][]{Ji2019, Fielding2020, Bustard2022}.

In fact, recent theoretical and numerical studies have increasingly emphasized the importance of turbulence in shaping the multiphase structure of galactic winds and the CGM. High-resolution simulations of cloud--wind interactions and turbulent mixing layers suggest that stochastic motions, shear-driven instabilities, and small-scale velocity fluctuations can significantly influence the survival, entrainment, and energetics of cool gas \citep[e.g.,][]{Ji2019, Fielding2020, Tan2021, Bustard2022, Chen2023, Hidalgo-Pineda2025}. In parallel, observational diagnostics such as velocity structure functions and spatially resolved spectroscopy have begun to probe the scale-dependent nature of CGM kinematics \citep[e.g.,][]{Chen2024, Warren2025}. Together, these developments suggest that turbulence may play a dynamically central role in galactic winds, motivating a reassessment of whether such systems are primarily outflow-dominated or instead regulated by stochastic motions. However, direct observational constraints on the relative importance of turbulence in multiphase galactic winds remain limited, largely due to the challenges of interpreting resonant UV line profiles with physically realistic CGM models.

In Paper~I, we addressed this challenge by introducing a three-dimensional Monte Carlo radiative transfer (RT) framework, \texttt{PEACOCK}, designed to jointly model \lya\ and multiple rest-frame ultraviolet metal lines within a clumpy, multiphase outflow geometry. By simultaneously fitting emission and absorption features across a broad range of ionization states, that study established a unified kinematic framework capable of reproducing the observed line profiles in a self-consistent manner. Applying this framework to 45 nearby star-forming galaxies from the CLASSY survey \citep{Berg2022}, supplemented by five additional starbursts with archival \textit{HST}/COS spectroscopy \citep{Heckman2015}, we derived constraints on clump covering factors, bulk outflow velocity profiles, microscopic and macroscopic turbulent velocities, clump covering factors, and ionic column densities. The joint multi-line analysis mitigated key degeneracies inherent in single-transition modeling and provided, for the first time, a coherent multiphase kinematic description of galactic winds across a broad ionization range.

The present work builds upon the modeling results of Paper~I to explore their broader physical implications. We will focus on the inferred kinematic and energetic properties of the CGM wind and examine the partition of stellar feedback energy between coherent bulk outflows and stochastic turbulent motions. We then investigate how these kinematic components scale with host galaxy properties such as stellar mass and star formation rate, and assess the dynamical picture of galactic winds that emerges when turbulence is treated as an energetically significant component. To this end, we analyze the relative contributions of bulk and turbulent motions to the kinetic energy and pressure budgets of the cool-to-warm CGM, evaluate how the inclusion of turbulence affects CGM--galaxy scaling relations, and examine the implications for the underlying feedback regime. We further characterize the velocity structure function of the inferred wind models in order to connect down-the-barrel spectroscopic constraints with turbulence diagnostics commonly employed in numerical simulations. Through this analysis, we assess whether galactic winds are best described as primarily outflow-dominated systems or as turbulence-regulated multiphase flows, and examine how feedback energy is distributed within the CGM.

The remainder of this paper is organized as follows. In Section~\ref{sec:data}, we briefly summarize the galaxy sample and the RT modeling results derived in Paper~I. Section~\ref{sec:CGM_connection} examines how the inferred kinematic and physical properties of the galactic winds connect to host galaxy properties. In Section~\ref{sec:turb_role}, we quantify the dynamical and energetic role of turbulence in galactic winds, including its contribution to the kinetic energy and pressure budgets. Section~\ref{sec:turb_drive} explores possible physical mechanisms responsible for the large turbulent velocities inferred in our analysis. In Section~\ref{sec:prev_works}, we compare our results with previous observational and theoretical studies. Section~\ref{sec:science_appl} discusses future applications of this framework to upcoming datasets and simulations. Finally, our main conclusions are summarized in Section~\ref{sec:conclusion}.

% %% The "ht!" tells LaTeX to put the figure "here" first, at the "top" next
% %% and to override the normal way of calculating a float position.
% %% The asterisk after "figure" tells the compiler to span multiple columns
% %% if a two column style is selected.
% \begin{figure*}[ht!]
% \plotone{AuthorChargeInfographic.png}
% \caption{The AAS journals are operated as a nonprofit venture, and author charges fairly recapture costs for the services provided in the publishing process. The chart above breaks down the services that author charges go toward. The AAS Journals' Business Model is outlined in a \href{https://aas.org/posts/news/2023/08/aas-open-access-publishing-model-open-transparent-and-fair}{2023 post}.
% \label{fig:general}}
% \end{figure*}

\section{Sample and RT Modeling Results}\label{sec:data}

This work builds directly upon the dataset described and the RT modeling presented in Paper~I. The underlying galaxy sample consists of 45 nearby star-forming galaxies drawn from the COS Legacy Archive Spectroscopic SurveY \citep[CLASSY;][]{Berg2022}, supplemented by five additional starburst systems from \citet{Heckman2015} to extend the dynamic range at high star formation rates. The data reduction, stellar continuum fitting, spectral resampling, and full details of the UV spectroscopic dataset are described in Paper~I and references therein.

In Paper~I, we applied the three-dimensional Monte Carlo RT framework \texttt{PEACOCK} to jointly model \lya\ and multiple rest-frame UV metal lines for each galaxy, deriving best-fit parameters describing the clumpy, multiphase structure and kinematics of their galactic winds. These inferred quantities include clump covering factors, bulk outflow velocity profiles, macroscopic turbulent velocity dispersions, and ionic column densities, among other galactic wind properties.

In the present work, we focus on analyzing the physical implications of the inferred RT modeling outputs. In particular, we investigate how the derived kinematic and energetic properties of the galactic wind correlate with host galaxy properties, quantify the role of turbulence in the wind energy budget, and examine the broader dynamical consequences for feedback-regulated galactic winds in the CGM.

\section{Connecting Galactic Wind Properties to Galaxy Properties}\label{sec:CGM_connection}

Understanding how the physical properties of the outflowing clumps in the galactic wind relate to the global properties of their host galaxies is essential for uncovering the physical mechanisms that regulate gas cycling around galaxies (e.g., \citealt{Tumlinson2017, Faucher2023}). In this section, we explore possible correlations between the clump properties derived from our RT modeling and the global galaxy properties, such as the total star formation rate (SFR) and stellar mass ($M_\star$). Our goal is to assess the extent to which the galactic wind properties trace (or are regulated by) the underlying galaxy properties.

\subsection{Ion Column Densities}

\begin{figure*}
\centering
\includegraphics[width=\textwidth]{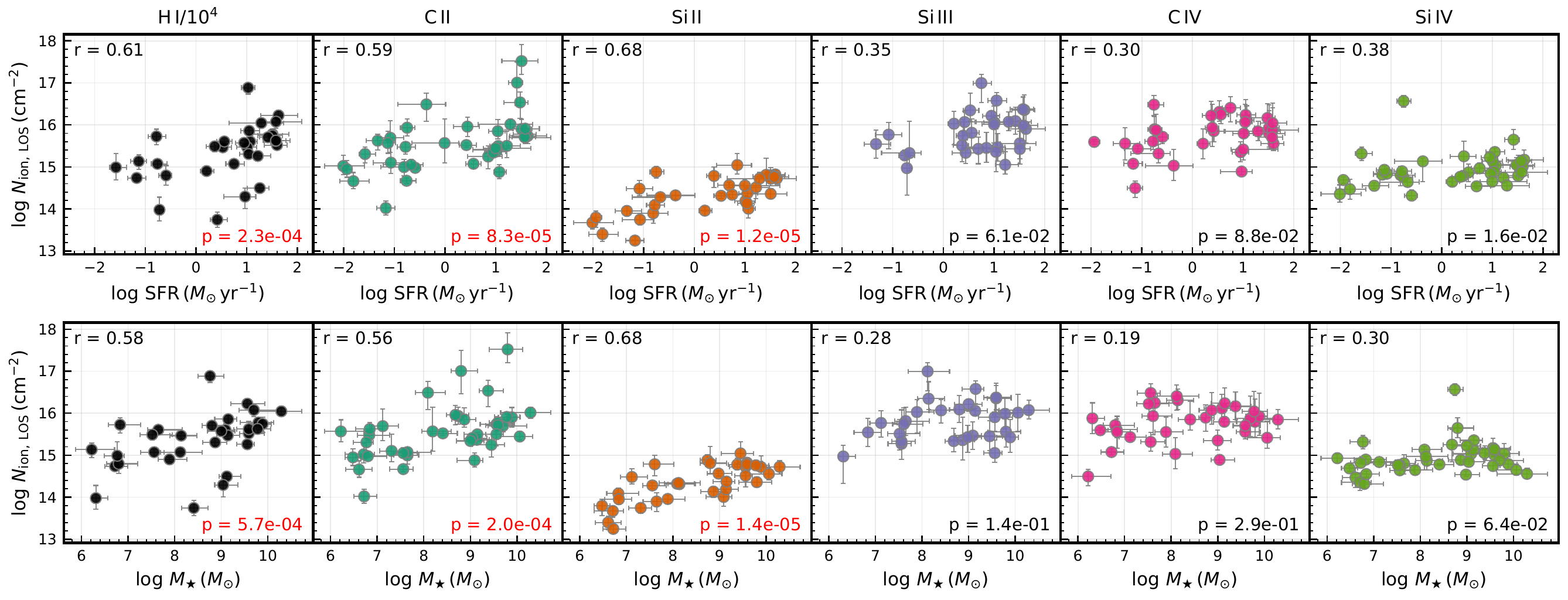}
    \caption{\textbf{Line-of-sight ionic column densities of clumps plotted as a function of global galaxy properties.} The top panels show $\log N_{\rm ion,\,LOS}$ versus total SFR, and the bottom panels show $\log N_{\rm ion,\,LOS}$ versus total $M_\star$, for six ions: \HI, \CII, \SiII, \SiIII, \CIV, and \SiIV. Each panel lists the Spearman rank correlation coefficient $r$ and $p$-value. All ions exhibit positive correlations with both SFR and $M_\star$, with \HI, \CII, and \SiII\ showing $>3\sigma$ significance, while Si\,\textsc{iii}, C\,\textsc{iv} and \SiIV\ display weaker trends. These results suggest that the cool, metal-enriched wind component scales with global galaxy growth, consistent with a picture in which stellar feedback enriches and maintains the gas in the CGM.
    \label{fig:ion_columns_sfr_mstar}}
\end{figure*}

Figure~\ref{fig:ion_columns_sfr_mstar} presents the relationships between the line-of-sight ion column densities and both the total SFR and $M_\star$ for six ions. All ions exhibit positive trends with both SFR and $M_\star$, although the statistical significance varies among species. Low-ionization tracers (\HI, \CII, and \SiII) show strong and statistically significant ($>3\sigma$) correlations with both SFR and $M_\star$, with \SiII\ exhibiting the tightest relation in both panels. In contrast, intermediate- and high-ionization species (\SiIII, \CIV, and \SiIV) display weaker trends that are not statistically significant.

The strong correlations of the low ions are consistent with the expectation that the amount of cool, metal-enriched gas in the wind increases with the intensity of ongoing star formation, as stellar feedback continuously enriches the outflowing material. The correlation with stellar mass likely reflects the deeper gravitational potential wells and higher integrated metal content of more massive galaxies, which can both retain baryons and accumulate larger reservoirs of enriched gas over time.

Notably, the similarity of the slopes and correlation strengths among the low-ion species suggests that these ions trace a common physical component of the wind. The weaker behavior of the higher ions may indicate that they arise from more extended, diffuse, or multiphase regions whose column densities are less directly coupled to global star formation or stellar mass.

\subsection{Turbulent and Outflow Velocities}

\begin{figure*}
\centering
\includegraphics[width=\textwidth]{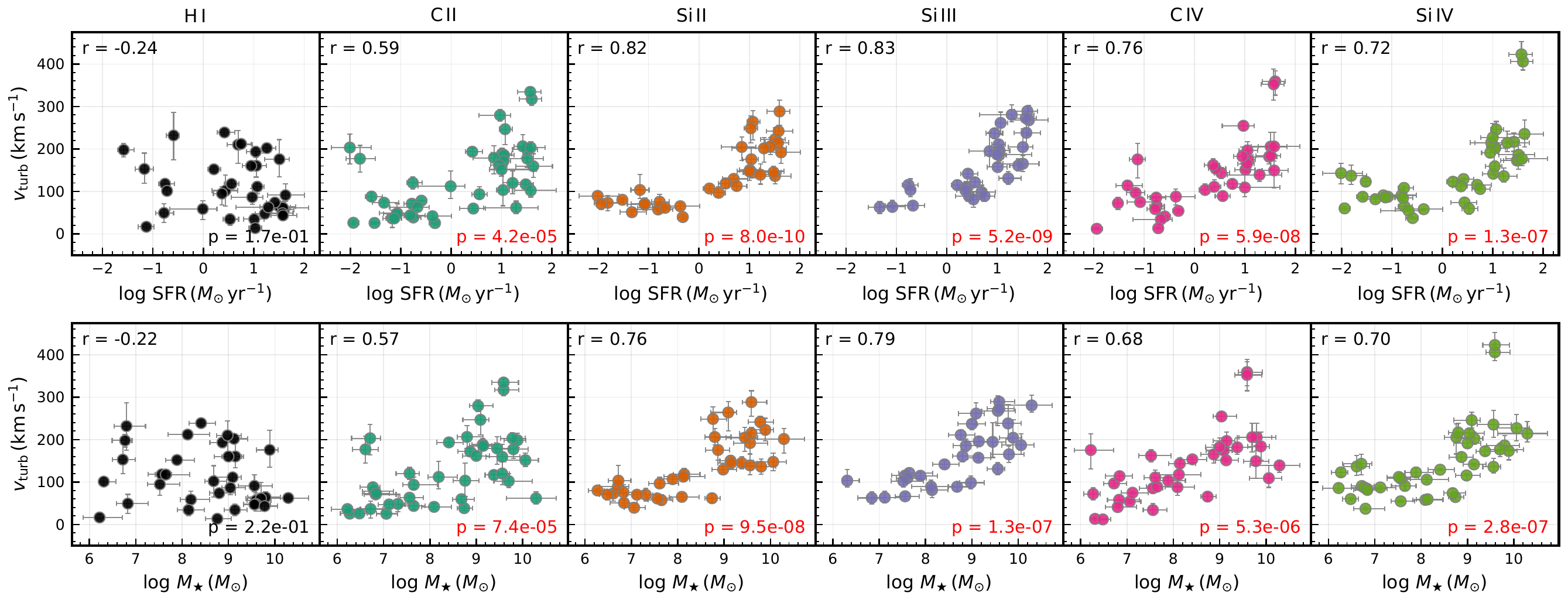}
    \caption{\textbf{Total turbulent velocity of clumps plotted as a function of global galaxy properties.} 
    The panels show the correlation between the inferred total turbulent velocity of individual clumps, $v_{\rm turb} = \sqrt{b_{\rm D,\,cl}^2 + \sigma_{\rm cl}^2}$, and both the total SFR (top row) and stellar mass (bottom row) for six ions tracing the cool to warm phases of the CGM. For all metal ions, we find a clear and statistically significant increase of $v_{\rm turb}$ with both SFR and $M_\star$, whereas \HI\ shows no significant correlation in either parameter space. The positive trends for the metal lines suggest that the dynamical state of the metal-enriched wind component scales with galaxy growth and star formation activity. In contrast, the absence of a similar trend in \HI\ may indicate that a substantial fraction of the neutral gas traces a more extended or ambient halo component whose dynamical state is less directly coupled to global galaxy properties.
    \label{fig:vturb_sfr_mstar}}
\end{figure*}

To investigate the origin of the complex velocity structure in the clumpy CGM wind, we examine how the total\footnote{We have also examined $b_{\rm D,\,cl}$ and $\sigma_{\rm cl}$ separately as functions of SFR and $M_\star$. Neither component alone shows correlations as significant as those observed for $v_{\rm turb}$, indicating that the total turbulent velocity provides a more robust tracer of the galaxy–wind dynamical connection.} turbulent velocity ($v_{\rm turb} = \sqrt{b_{\rm D,\,cl}^2 + \sigma_{\rm cl}^2}$) inferred from our RT modeling varies with global galaxy properties. Figure~\ref{fig:vturb_sfr_mstar} presents the relations between $v_{\rm turb}$ and both SFR and stellar mass for six ions probing the cool and warm phases.

We find a strong and statistically significant increase of $v_{\rm turb}$ with both SFR and $M_\star$ for all five metal ions, with Spearman coefficients $r \approx 0.5$ -- $0.85$ and $p \ll 10^{-4}$ in most cases. The only exception is \HI, which exhibits no significant correlation with either SFR or $M_\star$. These results suggest that the turbulent velocity of the metal-enriched CGM wind component scales with galaxy growth and star formation activity.

\begin{figure}
\centering
\includegraphics[width=0.48\textwidth]{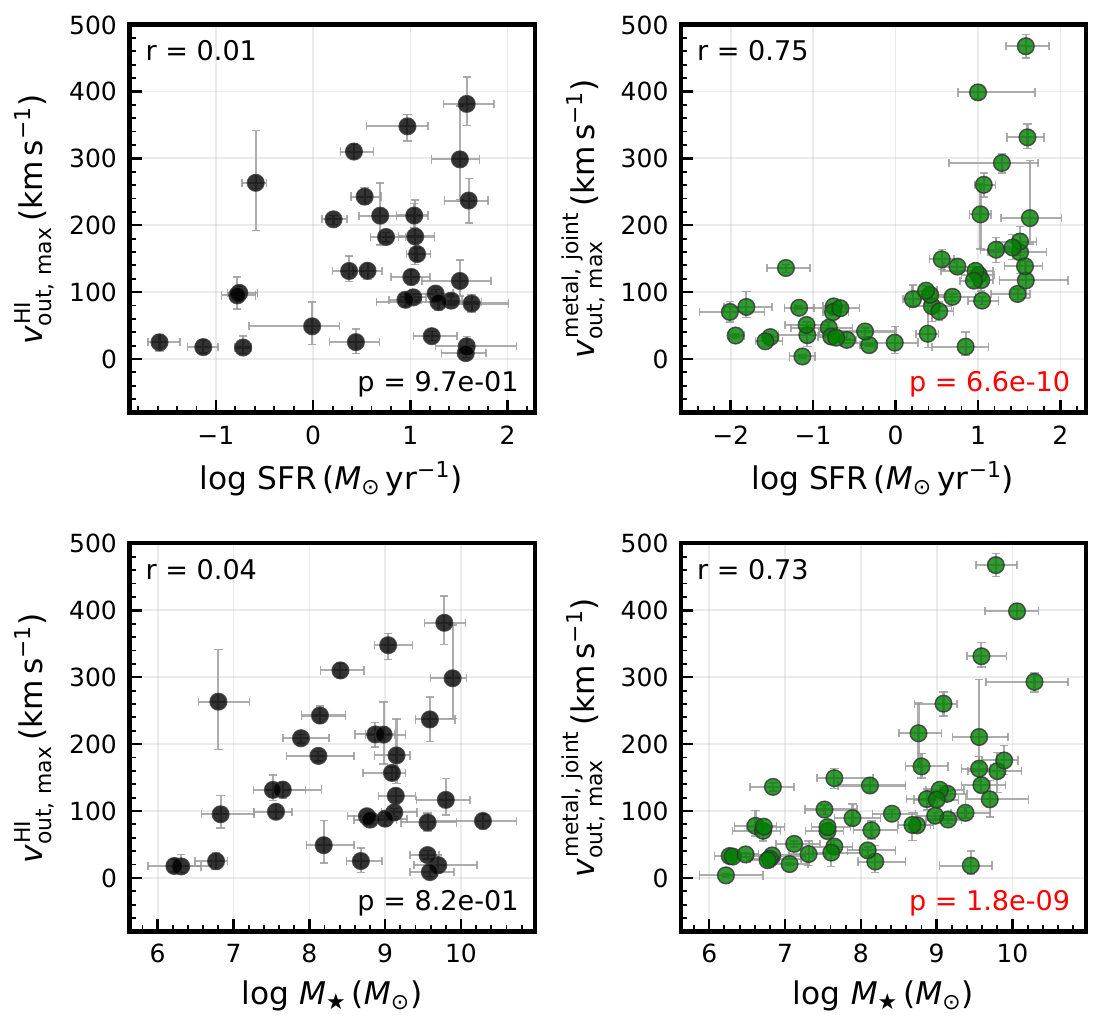}
    \caption{\textbf{Maximum outflow velocity of clumps as a function of global galaxy properties.} The figure shows the maximum outflow velocity $v_{\mathrm{out,\,max}}$ inferred from the best-fit RT models. \emph{Left panels:} maximum clump outflow velocities derived from \lya\ alone. \emph{Right panels:} maximum outflow velocities obtained from joint fitting of all available metal-line transitions. Pearson correlation coefficients $r$ and corresponding $p$-values are indicated in each panel. While $v_{\mathrm{out,\,max}}$ inferred from \lya\ alone shows no statistically significant correlation with either SFR or $M_\star$, the joint velocity exhibits strong and highly significant positive correlations with both properties, indicating that the bulk outflow velocity traced by metal lines more closely scales with global star formation activity and stellar mass than measurements based solely on \lya.
    \label{fig:vout_sfr_mstar}}
\end{figure}

In Figure~\ref{fig:vout_sfr_mstar}, we extend this analysis to the scaling of the clump maximum outflow velocity, $v_{\rm out,\,max}$. The left panels show the outflow velocity inferred from \HI\ based on \lya\ RT modeling, while the right panels present the results from our joint fitting of all available metal transitions for each galaxy. The multi-ion metal fits reveal highly significant positive correlations between $v_{\rm out,\,max}$ and both SFR and $M_\star$ ($r \approx 0.7$ – $0.75$, $p \ll 10^{-8}$), whereas the outflow velocity inferred from \HI\ alone shows no statistically significant correlation with either galaxy property. These findings indicate that the multi-ion outflow velocity provides a more robust tracer of the bulk kinetic energy of the metal-enriched wind component than measurements based solely on \HI.

Taken together, these trends support a physical picture in which both turbulence and bulk outflow velocities in the multiphase halo scale with the level of stellar feedback. Energy input from supernovae, radiation pressure, and stellar winds can inject kinetic energy into the CGM, enhancing turbulent motions and accelerating cool-to-warm gas to large velocities. The stronger correlations observed for metal ions relative to neutral hydrogen likely reflect differences in their spatial and thermal origins: metal ions predominantly trace gas that has been entrained, mixed, and enriched by recent feedback-driven outflows, whereas \HI\ may contain a substantial ambient halo component whose kinematics are less directly coupled to ongoing star formation.  The additional dependence on stellar mass further suggests that deeper gravitational potentials and longer integrated feedback histories in more massive galaxies help sustain higher turbulence levels and faster outflows over extended dynamical timescales.

\subsection{Ion Mass Outflow Rates}\label{sec:mass_outflow_rates}

Compared to velocity measurements alone, the mass outflow rate, $\dot{M}$, provides a more physically meaningful diagnostic of the efficiency with which feedback-driven winds redistribute baryons and metals from galaxies into their halos. We hereby compute the ion mass outflow rates for different species.

At radius $r$, the average number of clump intercepts per unit path length is $n_{\rm cl}(r)\sigma_{\rm cl}$. The ion mass flux through a sphere of radius $r$
is therefore
\begin{equation}
\dot{M}_{\rm ion}(r)
= 4\pi r^2\,v_{\rm out}(r)\, \bigl[m_{\rm ion}\,N_{\rm ion,\,cl}\,n_{\rm cl}(r)\sigma_{\rm cl}\bigr]
\label{eq:mdot_basic}
\end{equation}
where $m_{\rm ion}$ the mass per ion (atomic mass), $v_{\rm out}(r)$ the clump radial velocity at $r$, and $\sigma_{\rm cl}=\pi R_{\rm cl}^2$ is the geometric cross-section of a clump. Assuming a clump number density profile $n_{\rm cl}(r) = C r^{-2}$ and substituting the normalization constant $C$ from Eq. (6) of Paper~I, we obtain the compact expression for the ion mass outflow rate

\begin{figure*}
\centering
\includegraphics[width=\textwidth]{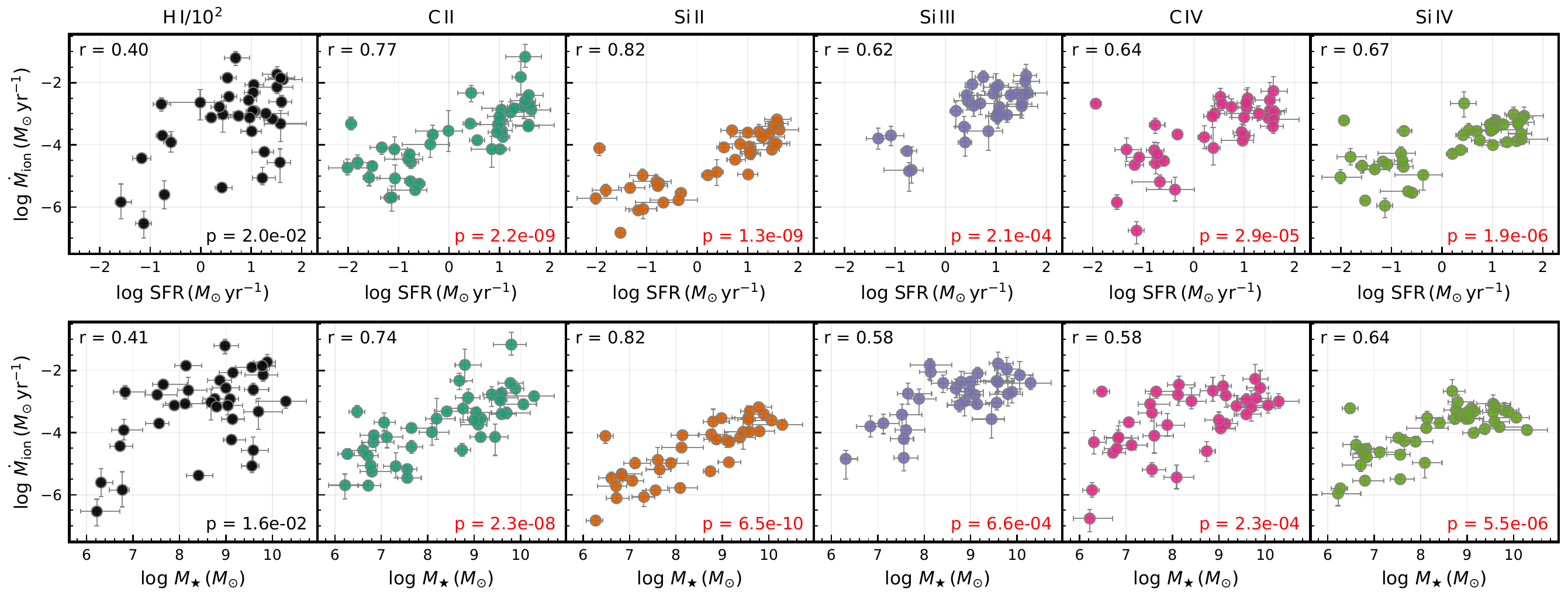}
    \caption{\textbf{Ion mass outflow rates ($\dot{M}_{\rm ion}$) as a function of global galaxy properties.} For \HI, we adopt the Ly$\alpha$-derived outflow velocity (scaled by $10^{-2}$ for visual clarity), while metal ions use the joint best-fit outflow velocity obtained from simultaneous modeling of all available transitions.  The metal species exhibit statistically significant positive correlations with both SFR and $M_\star$, whereas the \HI\ trend is weaker and displays larger scatter, primarily due to the dispersion in $v_{\rm out,\,max}$. The relative ordering $\dot{M}_{\rm C\,II} \simeq \dot{M}_{\rm C\,IV}$ and $\dot{M}_{\rm Si\,III} > \dot{M}_{\rm Si\,IV} > \dot{M}_{\rm Si\,II}$ mirrors the hierarchy observed in the line-of-sight column densities and suggests that a substantial fraction of the silicon mass resides in intermediate ionization states. Spearman rank coefficients $r$ and corresponding $p$-values are indicated in each panel.}
    \label{fig:mdot_sfr_mstar}
\end{figure*}

\begin{equation}
\dot{M}_{\rm ion}(r)
= 3\pi\,m_{\rm ion}\,N_{\rm ion,\,LOS}\,v_{\rm out}(r)\,
  \frac{R_{\rm in}R_{\rm out}}{R_{\rm out}-R_{\rm in}} \, 
\label{eq:mdot_compact}
\end{equation}
where $R_{\rm in}$ and $R_{\rm out}$ denote the inner and outer boundaries of the clump distribution, respectively. We adopt $R_{\rm in} = 2\,r_{50}$ as the clump launch radius, where $r_{50}$ is the galaxy half-light radius measured by \citet{Xu2022}, and assume $R_{\rm out} = 10\,R_{\rm in}$. Equation~(\ref{eq:mdot_compact}) shows that the radial variation of the ion mass outflow rate arises solely from the radial dependence of $v_{\rm out}(r)$. Physically, $\dot{M}_{\rm ion}$ scales with the outflow velocity $v_{\rm out}$, the total ionic column density along the line of sight, $N_{\rm ion,\,LOS}$, and the characteristic spatial extent of the clump distribution.

Figure~\ref{fig:mdot_sfr_mstar} shows how the ion mass outflow rates scale with both SFR and stellar mass for neutral hydrogen and five metal ions. In computing $\dot{M}_{\rm ion}$, we adopt the maximum clump velocity\footnote{See Appendix~\ref{app:radial_scaling} for examples of mass outflow rate profiles obtained with a radially varying $v_{\rm out}(r)$.} $v_{\rm cl,\,max}$ as a representative value of $v_{\rm out}(r)$. For \HI, we use the outflow velocity of the primary clump population inferred from \lya\ modeling (scaled by a factor of $10^{-2}$ for visual clarity), while all metal ions adopt the joint best-fit outflow velocity derived from simultaneous modeling of all available transitions.

All five metal ions exhibit strong and highly significant positive correlations between $\dot{M}_{\rm ion}$ and both SFR and $M_\star$, with Spearman coefficients $r \approx 0.5$ – $0.8$ and $p \ll 10^{-3}$ in most cases. \HI\ also shows a positive trend, but with substantially weaker significance and larger scatter compared to the metal species. The strength of these trends primarily reflects the combined impact of the scaling of both the ion column densities and the outflow velocities with galaxy properties. In particular, although $N_{\rm HI,\,LOS}$ shows strong correlations with SFR and stellar mass (see Figure~\ref{fig:ion_columns_sfr_mstar}), the large scatter in the \HI\ outflow velocity produces a much weaker dependence of $\dot{M}_{\rm HI}$ on host galaxy properties.

Another noteworthy feature is the relative ordering of the mass outflow rates among different ionization states of the same element. For carbon, we find $\dot{M}_{\rm C\,II} \simeq \dot{M}_{\rm C\,IV}$, whereas for silicon the ordering is strongly hierarchical, with $\dot{M}_{\rm Si\,III} > \dot{M}_{\rm Si\,IV} > \dot{M}_{\rm Si\,II}$, suggesting that a substantial fraction of the silicon mass resides in intermediate ionization states. A similar ordering is also present in the total line-of-sight column densities $N_{\rm ion,\,LOS}$ (see Figure \ref{fig:ion_columns_sfr_mstar}). 

\begin{figure}
\centering
\includegraphics[width=0.45\textwidth]{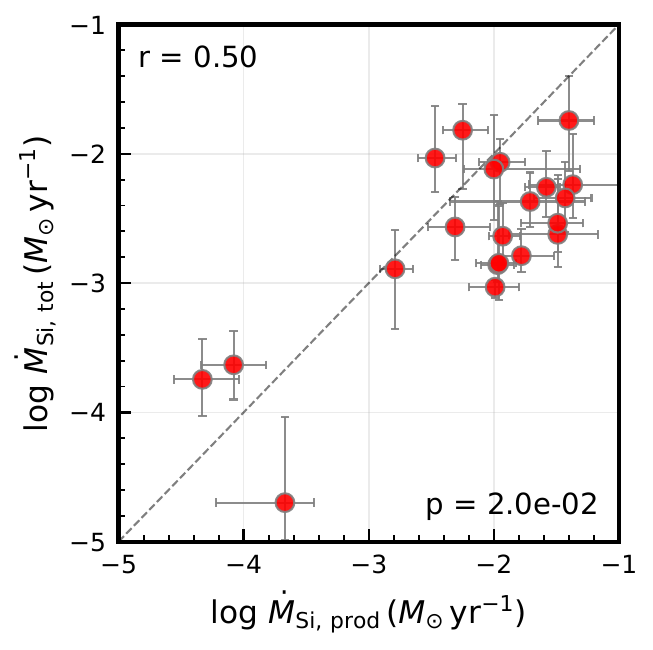}
    \caption{\textbf{Comparison between the total silicon mass outflow rate and the expected silicon production rate from star formation.} Only galaxies in which all three \SiII/\SiIII/\SiIV\ profiles are reliably measured are included. The expected silicon production rate, $\dot{M}_{\mathrm{Si,\,prod}}=10^{-3}\,\mathrm{SFR}$, is inferred from \textsc{Starburst99} models. A moderate positive correlation is observed, indicating that the cool-to-warm silicon mass flux broadly scales with the nucleosynthetic output from star formation. Most galaxies lie below the one-to-one relation (dashed line), suggesting that only a fraction of the freshly synthesized silicon is present in the cool-to-warm phase traced by these ions at a given time. This offset likely reflects incomplete entrainment of metals into the cool-to-warm component of the outflow, with the remaining silicon residing in other wind phases (e.g., highly ionized hot gas), being depleted due to dust extinction, or being recycled back into the ISM.}
    \label{fig:mdot_Si_compare}
\end{figure}

To provide a quantitative comparison with the expected silicon output from stellar feedback, we estimate the total silicon mass outflow rate as the sum of the three observed ion stages, $\dot{M}_{\rm Si,\,tot} = \dot{M}_{\rm Si\,II} + \dot{M}_{\rm Si\,III} + \dot{M}_{\rm Si\,IV}$, and compare it with the silicon production rate inferred from the SFR. We report values of $\dot{M}_{\rm Si}$ only for galaxies in which all three silicon transitions are reliably detected and their line profiles are well measured. Adopting \textsc{Starburst99} metal yields (\citealt{Leitherer1999}; see also \citealt{Heckman2015}), we assume a characteristic silicon production rate of $\dot{M}_{\rm Si,\,prod} = 10^{-3}\,\mathrm{SFR}$, which provides a useful benchmark for assessing the efficiency of metal entrainment in the observed outflows.

Figure~\ref{fig:mdot_Si_compare} shows a moderate positive correlation between $\dot{M}_{\rm Si,\,tot}$ and $\dot{M}_{\rm Si,\,prod}$ ($r=0.49$, $p=2.7\times10^{-2}$), broadly consistent with previous findings (e.g., \citealt{Xu2022}). 
The data span both sides of the one-to-one relation, although the majority of galaxies lie systematically below it, indicating that the silicon mass flux traced by \SiII, \SiIII, and \SiIV\ is typically smaller than the silicon production rate.

This offset suggests that only a fraction of the newly synthesized silicon is present in the cool-to-warm clumpy phase probed by \SiII, \SiIII\ and \SiIV\ at a given time. Several physical effects may contribute to this difference. First, only part of the total metal yield may be efficiently entrained into the observable cool-to-warm clumps, while the remainder resides in other phases of the multiphase wind, such as highly ionized hot outflows, metal-rich supernova ejecta temporarily confined within the ISM, or metals locked in stellar remnants.  Second, a portion of the transported silicon may be depleted onto dust grains during its passage through the ISM and CGM, thereby reducing the gas-phase silicon detectable through UV absorption. Both effects would act to lower $\dot{M}_{\rm Si,\,tot}$ relative to the silicon production rate. Conversely, in systems where $\dot{M}_{\rm Si,\,tot}$ exceeds $\dot{M}_{\rm Si,\,prod}$, the outflow may be dominated by the acceleration and entrainment of pre-enriched CGM material rather than solely freshly synthesized ejecta from the starburst. In this scenario, the silicon carried by the wind represents a mixture of newly produced metals and pre-existing silicon originally residing in the CGM clouds.

Overall, the positive correlations between $\dot{M}_{\rm ion}$ and both SFR and $M_\star$ indicate that the mass flux of metal-enriched cool-to-warm gas scales with global galaxy growth and star formation activity. In this framework, the observed ion mass outflow rates trace the fraction of the total metal yield that is entrained into the cool-to-warm clumpy phase of the wind and distributed among gas with different ionization states. The remaining metals are likely carried by other phases of the feedback-driven flow, including highly ionized hot winds (e.g., traced by X-ray–emitting plasma), depleted from the gas phase onto dust grains, or recycled back into the ISM before reaching large radii. The systematic offsets below the one-to-one relation between $\dot{M}_{\rm Si,\,tot}$ and $\dot{M}_{\rm Si,\,prod}$ therefore suggest that metal production and metal transport are not perfectly coupled within the observable cool-to-warm phase. Future constraints on higher-ionization tracers and on the hot-gas component will be essential for closing the metal budget and establishing a more complete picture of the multiphase structure and energetics of galactic winds.

\subsection{Total Mass Outflow Rates and Mass-Loading Factors}\label{sec:mdot_eta}

The total gas mass outflow rate, $\dot{M}_{\rm gas}$, and the corresponding mass-loading factor $\eta$, provide a direct measure of how efficiently star formation couples to the circumgalactic gas. These quantities therefore provide a more physically comprehensive probe of feedback processes than mass outflow rates inferred from individual ionic species alone.

We hereby estimate the total hydrogen mass outflow rate by first summing the silicon mass flux traced by the low- and intermediate-ionization stages, and then converting the resulting silicon mass flux into a corresponding hydrogen mass flux using the galaxy metallicity. Specifically, we compute the mass outflow rates of \SiII, \SiIII, and \SiIV\ following the methodology described in Section~\ref{sec:mass_outflow_rates}, and define the total silicon mass outflow rate as the sum of these three components:

\begin{equation}
\dot{M}_{\rm Si} = \dot{M}_{\rm Si\,II} + \dot{M}_{\rm Si\,III} + \dot{M}_{\rm Si\,IV}
\label{eq:mdot_Si}
\end{equation}

We then infer the gas-phase metallicity relative to solar from the nebular oxygen abundance via
\begin{equation}
\frac{Z_{\rm gal}}{Z_\odot} = 10^{\left[(12+\log({\rm O/H})) - (12+\log({\rm O/H}))_\odot\right]}
\end{equation}
adopting a solar oxygen abundance of $(12+\log({\rm O/H}))_\odot = 8.69$. 
Assuming that silicon scales linearly with the total metallicity, and adopting a solar silicon mass fraction $Z_{{\rm Si},\,\odot}$ \citep{Asplund2021}, the total hydrogen mass outflow rate is then given by (see also \citealt{Huberty2024})
\begin{equation}
\dot{M}_{\rm H} \;=\; \frac{\dot{M}_{\rm Si}}{Z_{\rm gal}\, Z_{{\rm Si},\,\odot}}
\label{eq:mdot_H}
\end{equation}

To facilitate direct comparison with theoretical feedback models, which generally report total gas mass outflow rates including helium, we convert the hydrogen mass outflow rate to a total gas mass outflow rate via

\begin{equation}
\dot{M}_{\rm gas} = 1.36\dot{M}_{\rm H}
\label{eq:mdot_gas}
\end{equation}
where the factor of 1.36 accounts for the contribution of helium assuming a primordial helium mass fraction. Throughout this work, unless otherwise specified, we define the mass-loading factor using the total gas mass outflow rate

\begin{equation}
\eta \equiv \frac{\dot{M}_{\rm gas}}{{\rm SFR}}
\end{equation}

\begin{figure*}
\centering
\includegraphics[width=0.9\textwidth]{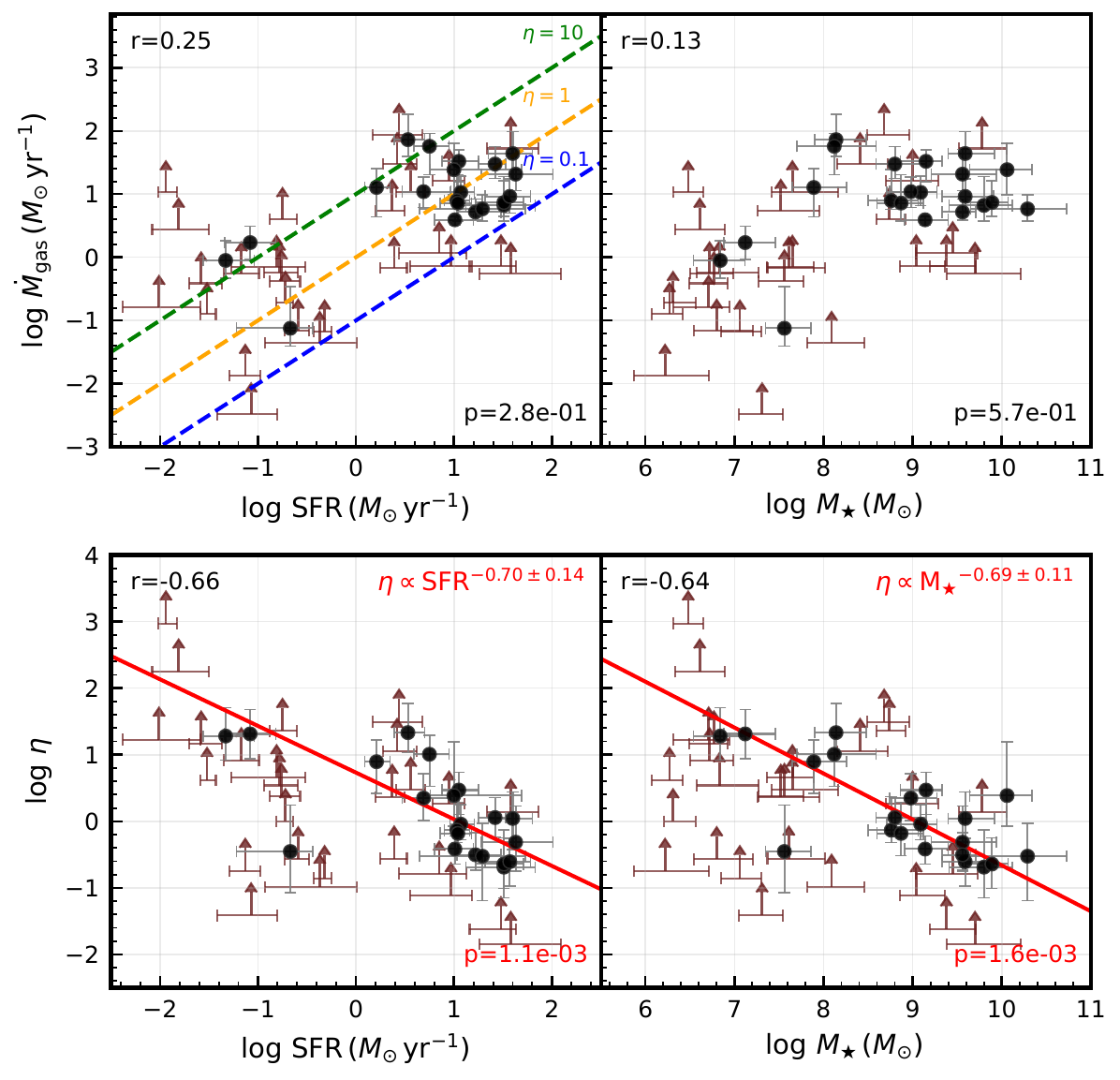}
    \caption{\textbf{Scaling relations between the total gas mass outflow rate, mass-loading factor, star formation rate, and stellar mass.} The top row shows the total gas mass outflow rate, $\dot{M}_{\rm gas}$, as a function of SFR (left) and stellar mass $M_\star$ (right). Black circles with error bars denote galaxies with measurements of all three silicon ions (\SiII, \SiIII, and \SiIV), while dark red upward arrows indicate lower limits when one or more ionic components are not detected. Colored dashed lines indicate constant mass-loading factors of $\eta = \dot{M}_{\rm gas}/{\rm SFR} = 0.1$, 1, and 10. A mild positive correlation is observed between $\dot{M}_{\rm gas}$ and SFR, while no significant trend is found with stellar mass. The bottom row shows the corresponding mass-loading factor, $\eta$, as a function of SFR (left) and stellar mass (right). Linear regressions are carried out using only data points with uncertainties in both axes, with errors in both $x$ and $y$ incorporated into the fitting procedure. Strong anti-correlations are found between $\eta$ and both SFR and stellar mass, and the best-fit power-law slopes strongly favor an energy-driven feedback scenario.
    \label{fig:eta_v}}
\end{figure*}

The top row of Figure~\ref{fig:eta_v} shows the scaling of the total gas mass outflow rate with star formation rate and stellar mass. By construction, the inferred mass outflow rate scales as $\dot{M}_{\rm gas}\propto N_{\rm gas,\,LOS}\,v_{\rm out}$. Galaxies with detections of all three silicon ionic species (\SiII, \SiIII, and \SiIV) are shown as black points with error bars, while systems lacking one or more ionic components are represented as lower limits (upward arrows). We find a mild positive correlation between $\dot{M}_{\rm gas}$ and SFR ($r = 0.26$), while no statistically significant trend is observed with stellar mass ($r = 0.14$). Across the sample, the majority of galaxies exhibit mass-loading factors in the range $0.1 \lesssim \eta \lesssim 10$, broadly consistent with previous CGM studies (e.g., \citealt{Heckman2015}).

The bottom row of Figure~\ref{fig:eta_v} shows the corresponding scaling of the mass-loading factor $\eta$ with star formation rate and stellar mass. The mass-loading factor quantifies the efficiency with which star formation-driven feedback expels gas relative to the rate at which new stars form, thereby directly linking the baryon cycle to feedback energetics. We find a strong ($>3\sigma$) anti-correlation between $\eta$ and both SFR and stellar mass, indicating that galaxies with higher SFR and stellar mass tend to drive outflows with lower mass-loading efficiencies. 

In particular, the scaling with SFR is well described by $\eta \propto {\rm SFR}^{-0.70 \pm 0.15}$, with a similarly strong anti-correlation is with stellar mass, $\eta \propto M_\star^{-0.69 \pm 0.11}$. Such a scaling with SFR is remarkably consistent with expectations for energy-driven outflows. In an energy-conserving scenario, the injected feedback energy scales with star formation rate such that $\dot{M_{\rm gas}}\,v_{\rm out}^{2} \propto {\rm SFR}$. Given that in our model the inferred mass outflow rate scales linearly with velocity, i.e., $\dot{M}_{\rm gas} \propto v_{\rm out}$, this leads to $\eta \propto {\rm SFR}^{-2/3}$. By contrast, momentum-driven winds which satisfy $\dot{M} v \propto {\rm SFR}$, predict a significantly shallower dependence, $\eta \propto {\rm SFR}^{-1/3}$. The observed slope therefore favors an energy-driven interpretation.

Taken together, these results support a physical picture in which the mass-loading efficiency of galactic outflows is systematically regulated by galaxy growth. The observed scalings with star formation rate and stellar mass are broadly consistent with expectations for energy-driven feedback, suggesting that the coupling between stellar energy injection and large-scale gas flows remains dynamically important. The resulting mass-loading relation therefore reflects the underlying energetics of star formation–driven winds.

\section{Quantifying the Critical Role of Turbulence in Galactic Winds}\label{sec:turb_role}

One of the key innovations of this study is the disentanglement of turbulent velocity from the bulk outflow motion in the multiphase, clumpy galactic wind. While observational analyses of galaxy winds have traditionally focused on bulk outflow velocities inferred from line centroids or maximum absorption depths, the turbulent velocity component has received comparatively less direct attention (see, however, recent studies of CGM velocity structure functions by \citealt{HWChen2023, Chen2024, Chen2025}).

In many analyses, turbulence is treated primarily as a nuisance parameter for line broadening rather than as a physically meaningful carrier of kinetic energy. However, in a multiphase CGM halo shaped by star formation feedback, turbulence is more than a source of stochastic broadening: it represents a key channel through which feedback energy cascades from galactic scales into small-scale motions, influences the survival and entrainment of cool clumps, and regulates phase exchange through mixing, conduction, and dissipation (e.g., \citealt{Faucher2023, Gronke2026}).

In this section, we quantify the role of turbulence in the galactic wind from several complementary perspectives and demonstrate that it constitutes an essential component of wind energetics that merits explicit consideration alongside bulk outflow motions.

\subsection{Turbulence as a Major Kinematic Component}\label{sec:turb_dominates}

Since the turbulent velocity directly contributes to the kinetic energy budget of individual clumps, it provides a complementary diagnostic to the bulk outflow velocity. To quantify the relative importance of turbulence and coherent outflow, we introduce the dimensionless parameter
\begin{equation}
    R_{\sigma/v} = \frac{v_{\rm turb}}{v_{\rm out}}
\end{equation}
defined as the ratio of clump total turbulent velocity to clump (maximum) radial outflow velocity. This ratio characterizes whether the clump kinematics are dominated by coherent bulk motion or by turbulent motions. In the regime $R_{\sigma/v} \ll 1$, the bulk outflow velocity exceeds the turbulent velocity, and the large-scale outflow sets the dominant kinematic scale. Conversely, when $R_{\sigma/v} \gtrsim 1$, turbulent motions contribute comparably to or exceed the bulk velocity, indicating a qualitatively distinct regime in which the clump dynamics are no longer governed by large-scale wind acceleration but instead by other processes, such as shear instabilities in the mixing layers, turbulent pressure support, and local energy dissipation.

\begin{figure*}
\centering
\includegraphics[width=\textwidth]{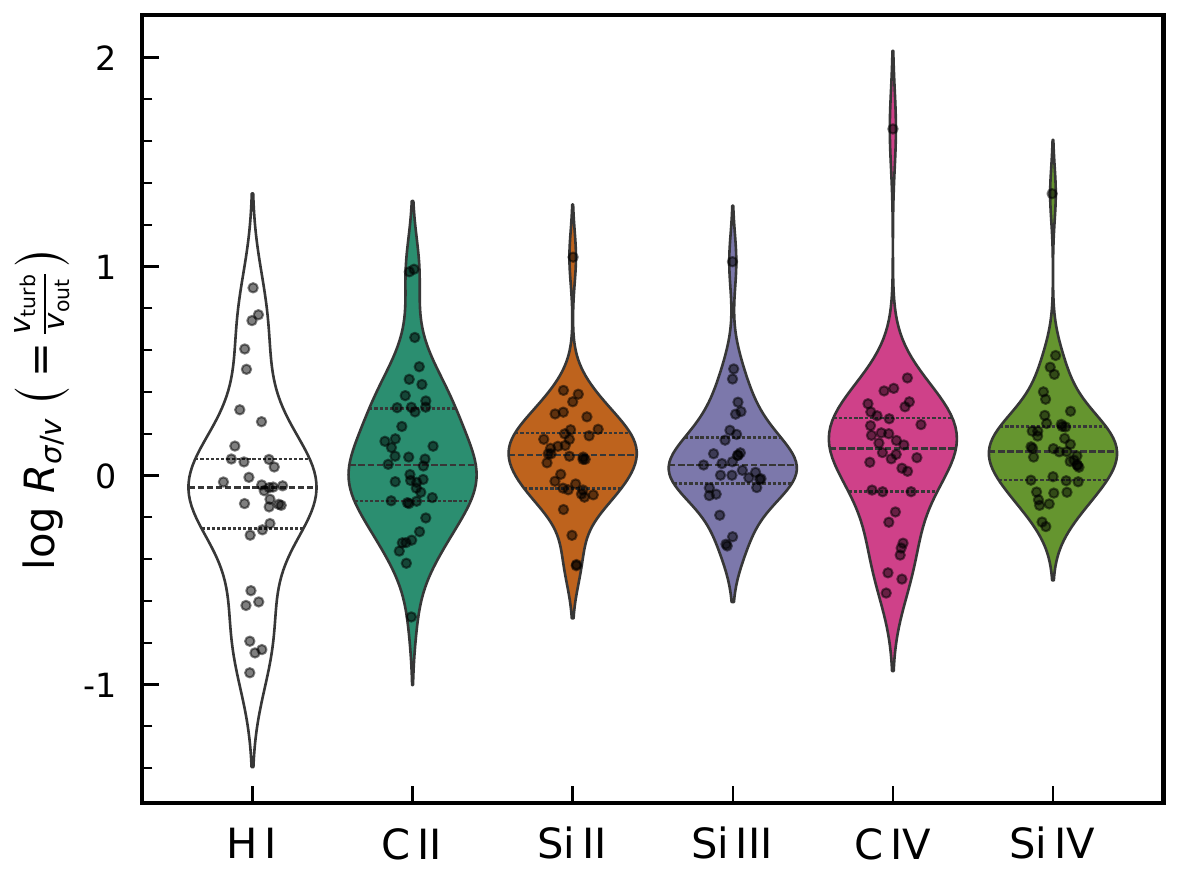}
    \caption{\textbf{Distributions of the ratio between total turbulent velocities and bulk outflow velocities for all six ions analyzed in this work.} Each violin shows the distribution of $v_{\rm turb}/v_{\rm out}$ across the galaxy sample, with gray points indicating measurements for individual galaxies. While only about one-third of \HI-traced clumps have $v_{\rm turb} > v_{\rm out}$, approximately half to three-quarters of the metal-traced systems exhibit turbulent velocities comparable to or exceeding the coherent outflow speed. These results indicate that turbulent motions constitute a dynamically significant component of the wind kinematics, rather than acting as a minor broadening term superimposed on a dominant bulk flow.
    \label{fig:vturb_vout_violin}}
\end{figure*}
Figure~\ref{fig:vturb_vout_violin} shows the distribution of $\log R_{\sigma/v}$ for all six transitions. Each violin represents the distribution across our galaxy sample, with individual points corresponding to measurements for each galaxy. Across all ions, turbulence is generally significant relative to the bulk outflow velocity: the distributions of $\log R_{\sigma/v}$ are broad (spanning $\sim$\,1 -- 2 dex) and are typically centered near $R_{\sigma/v}\sim 1$. Quantitatively, the fractions of galaxies with $\log R_{\sigma/v}>0$ (i.e.\ $v_{\rm turb} > v_{\rm out}$) are
\[
f(\log R_{\sigma/v}>0)= \{35\%, 55\%, 67\%, 65\%, 69\%, 71\%\}
\]
for \HI, \CII, \SiII, \SiIII, \CIV, and \SiIV, respectively. Thus, only about one third of the \HI-traced clumps exhibit turbulent velocities that are larger than the coherent outflow velocity, whereas roughly half to two thirds of the metal-traced clumps --- particularly the intermediate and high ions --- satisfy $v_{\rm turb} \gtrsim v_{\rm out}$. 

This systematic increase from neutral to metal ions indicates that metal-enriched gas preferentially occupies a regime in which turbulent motions are dynamically comparable to, or even exceed, the coherent bulk outflow. Across all metal ions, the prevalence of $R_{\sigma/v}\sim 1$ -- 10 demonstrates that turbulence cannot be treated as a small perturbation to the kinematics of the galactic wind, but instead constitutes a major component of the kinetic energy budget in the multiphase halo.

We find this result particularly intriguing, as it motivates a refinement of the conventional picture of galactic winds. Traditionally, galactic outflows are described as large-scale, coherent flows accelerated by stellar and supernova feedback, with the bulk motion assumed to carry most of the injected kinetic energy (e.g., \citealt{Chevalier1985, Veilleux2005, Heckman2015}). In contrast, our results indicate that a substantial fraction of the kinetic energy is instead manifested as turbulent motions both within and among the clumps, and that the total turbulent velocity often equals or exceeds the coherent outflow speed. This implies that turbulence is not merely a secondary broadening component superimposed on a dominant bulk flow, but rather a major channel through which feedback energy is redistributed within the multiphase halo. Turbulent motions mediate the exchange of momentum and thermal energy between gas phases, regulate mixing and phase transformation, and contribute significantly to the overall kinetic energy budget of the galactic wind.

\subsection{Macroscopic Turbulence as the Leading Contributor to the Kinematic Pressure}\label{sec:macro_dominates_p}

\begin{figure*}
\centering
\includegraphics[width=0.9\textwidth]{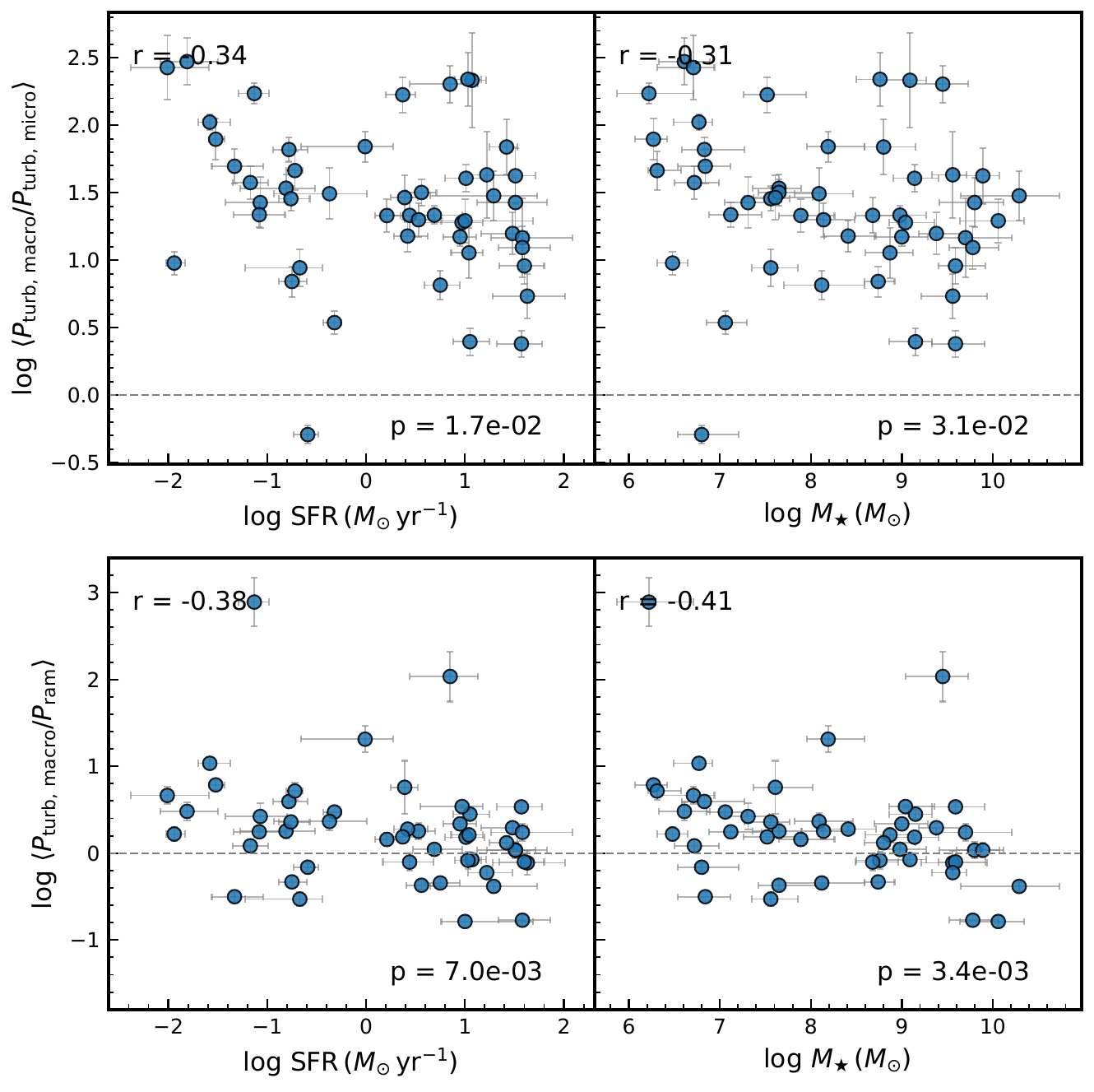}
    \caption{\textbf{Comparison of the average pressure ratios $\langle P_{\rm turb,macro}/P_{\rm turb,micro} \rangle$ (top panels) and $\langle P_{\rm turb,macro}/P_{\rm ram} \rangle$ (bottom panels) as functions of SFR (left) and stellar mass $M_\star$ (right).} All quantities are shown in logarithmic units. The dashed horizontal line marks unity, where macroscopic turbulent pressure equals either microscopic turbulence or ram pressure. Both ratios exhibit moderate anti-correlations with SFR and $M_\star$ ($\sim$\,2.2 -- 2.5$\sigma$ significance), indicating a mild decline of the ratios toward higher-mass and higher-SFR systems. Nevertheless, the ratios remain predominantly above unity across the full dynamic range, demonstrating that macroscopic turbulence typically dominates over both internal clump turbulence and coherent outflow in setting the kinematic pressure of the cool-to-warm gas.}
    \label{fig:Pturb_macro_micro}
\end{figure*}

We now quantify the relative importance of three components of clump motion in the cool-warm phase: 
macroscopic turbulence, microscopic turbulence, and bulk outflow. These components are characterized by the random velocity dispersion among clumps ($\sigma_{\rm cl}$), the clump Doppler parameter ($b_{\rm D,\,cl}$), and the clump radial outflow velocity ($v_{\rm out}$), respectively. We evaluate their relative roles by examining their contributions to the kinematic pressure of the cool gas, which -- as we will show below -- is proportional to the second moment of the clump velocity distribution.

The microscopic turbulent pressure is characterized by the Doppler parameter $b_{\rm D,\,cl}$, which reflects the internal thermal and non-thermal motions within individual clumps. The macroscopic turbulent pressure, however, corresponds to the random motion among the clumps. Therefore, the relative importance of macroscopic to microscopic turbulent support scales as 

\begin{equation}
\frac{P_{\rm turb,\,macro}}{P_{\rm turb,\,micro}}
\sim
\left(\frac{\sigma_{\rm cl}}{b_{\rm D,\,cl}}\right)^2
\end{equation}

Similarly, the ratio between macroscopic turbulent pressure and ram pressure associated with the bulk outflow scales as

\begin{equation}
\frac{P_{\rm turb,\,macro}}{P_{\rm ram}}
\sim
\left(\frac{\sigma_{\rm cl}}{v_{\rm out}}\right)^2
\end{equation}

Thus, the dynamical hierarchy among the three pressure components is determined by the squared ratios of their characteristic velocities. For each transition, we compute these ratios and take the mean across all available metal lines for each galaxy, obtaining 
\(\langle P_{\rm turb,\,macro}/P_{\rm turb,\,micro}\rangle\) and \(\langle P_{\rm turb,\,macro}/P_{\rm ram}\rangle\).

In Figure~\ref{fig:Pturb_macro_micro}, we compare the average ratios 
$\langle P_{\rm turb,\,macro}/P_{\rm turb,\,micro}\rangle$ and 
$\langle P_{\rm turb,\,macro}/P_{\rm ram}\rangle$ as functions of SFR and stellar mass. We find that $\langle P_{\rm turb,\,macro}/P_{\rm turb,\,micro}\rangle$ 
typically lies in the range $\sim 10$ -- $100$ for most galaxies. Since the clump Doppler parameter includes both thermal and non-thermal contributions, this implies the ordering
\[
\sigma_{\rm cl} \gg b_{\rm D,\,cl} > b_{\rm th,\,cl}
\]
indicating that macroscopic clump-to-clump motions dominate over internal turbulent and thermal broadening within the cool-to-warm gas phase.

Likewise, for most galaxies we find 
$\langle P_{\rm turb,\,macro}/P_{\rm ram}\rangle \gtrsim 1$, implying
\[
\sigma_{\rm cl} \gtrsim v_{\rm out}
\]
i.e., macroscopic turbulent motions are generally comparable to or exceed the bulk outflow velocity in setting the kinetic support of the gas. While both ratios exhibit moderately significant anti-correlations with SFR and $M_\star$ (Spearman $r \sim -0.3$ to $-0.4$, with $p$ values corresponding to $\sim$\,2.2 -- 2.5$\sigma$ significance), the scatter at fixed SFR or $M_\star$ remains large.

Taken together, this pressure decomposition indicates that macroscopic turbulent motions are typically the primary contributor to the total kinematic pressure of the cool-to-warm gas, exceeding both microscopic turbulence and ram pressure in most systems. The modest decline of the pressure ratios with increasing SFR and stellar mass suggests that galaxies with stronger global activity may exhibit slightly enhanced bulk coherence relative to random clump motions. Nevertheless, the persistence of macroscopic turbulence as the primary support mechanism across the full dynamical range points toward an approximately self-similar kinematic structure of galactic winds in the CGM. This behavior implies that feedback energy injected by star formation is redistributed across multiple physical scales in a way that preserves a roughly constant partition between large-scale turbulence, internal clump turbulence, and coherent outflow. Such scaling is consistent with a self-regulated cascade of kinetic energy from global outflows to small-scale turbulent motions within the multiphase medium.

\subsection{Turbulence Strengthens CGM–Galaxy Scaling Relations}\label{sec:turb_dominates_scaling}

In studies of galactic winds, scaling relations are most commonly constructed using the bulk outflow velocity, under the assumption that the coherent wind speed captures the dominant kinetic response of the CGM to star formation feedback. However, our analysis in previous sections shows that a substantial fraction of kinetic energy in the cool-warm gas resides in turbulent motions rather than in the coherent bulk outflow. If turbulence constitutes a major energy reservoir, then scaling relations based solely on the bulk outflow velocity may provide an incomplete description of the feedback imprint on the galactic winds in the CGM.

\begin{figure*}
\centering
\includegraphics[width=\textwidth]{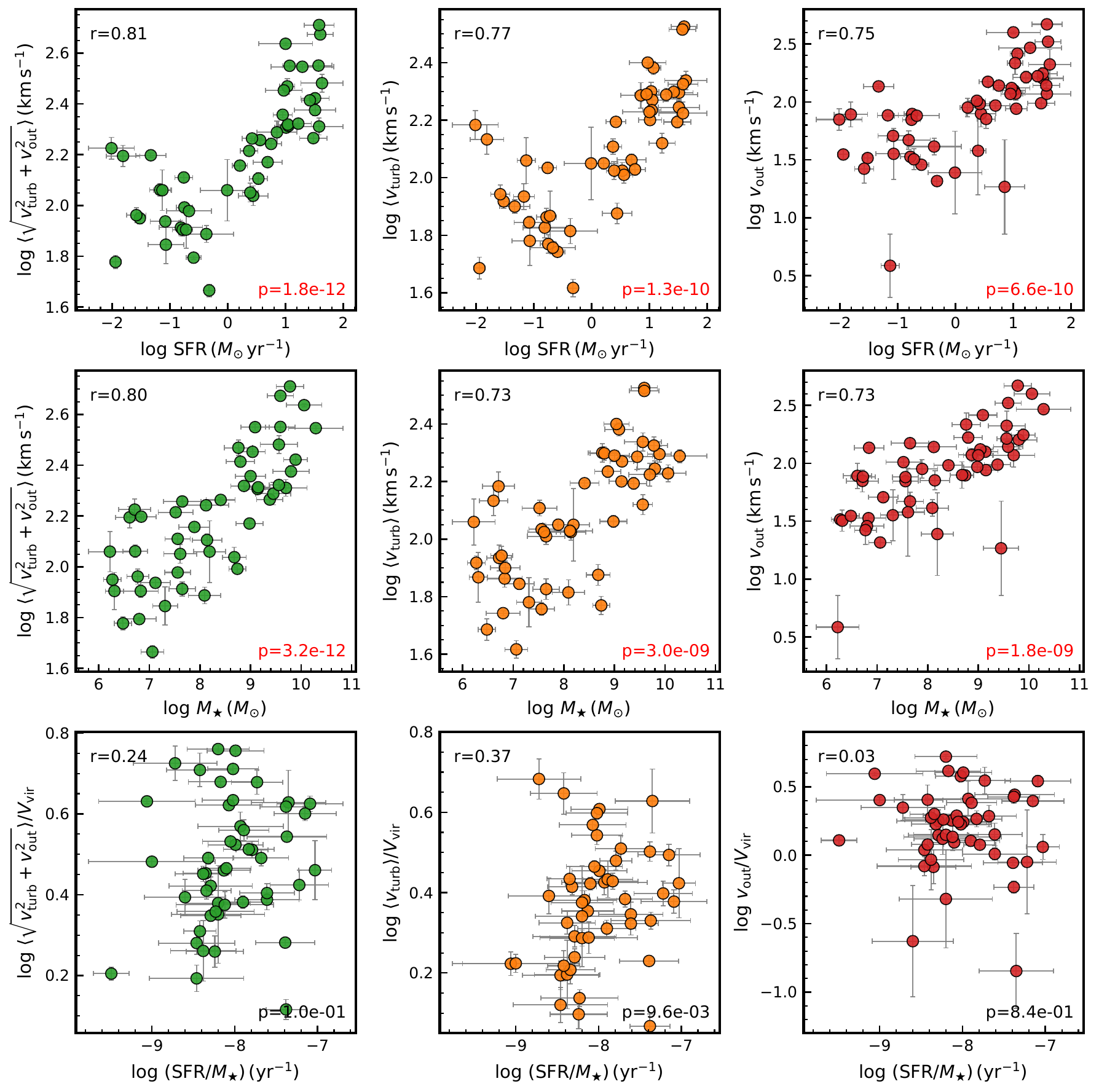}
    \caption{\textbf{Scaling relations between the metal–ion–averaged CGM wind velocities and global galaxy properties.} The three characteristic velocities are: total velocity $v_{\rm tot} = \sqrt{v_{\rm turb}^2 + v_{\rm out}^2}$ (green), turbulent velocity (orange), and bulk outflow velocity (red). The top and middle rows plot velocities versus $\rm SFR$ and $M_\star$, while the bottom row shows velocities normalized by the halo virial velocity $V_{\rm vir}$ versus the specific SFR. Each panel lists the Spearman $r$ and the associated $p$ value. All three velocity measures exhibit strong positive correlations with SFR and $M_\star$, with the tightest relations obtained for the total velocity. This indicates that the combined contribution of turbulent and coherent motions more closely traces the cumulative impact of star formation feedback than either component alone. After normalizing by $V_{\rm vir}$, the velocity–sSFR relations become substantially weaker. The normalized velocities exhibit at most tentative positive trends with sSFR and at reduced statistical significance, indicating that CGM wind kinematics are more strongly regulated by the overall energy scale set by star formation and galaxy mass than by the star formation efficiency.
    \label{fig:vtot_SFR_Mstar}}
\end{figure*}

We hereby compare how the characteristic velocities of the cool–warm CGM scale with global galaxy properties when turbulence is included or excluded from the effective velocity definition. Figure~\ref{fig:vtot_SFR_Mstar} presents these scaling relations using three different velocity measures:
(1) the total kinetic velocity, $v_{\rm tot} = \sqrt{v_{\rm turb}^2 + v_{\rm out}^2}$;
(2) the turbulent velocity alone, $v_{\rm turb}$; and
(3) the bulk outflow velocity alone, $v_{\rm out}$. All velocities represent metal–ion–averaged values for each galaxy.

All three velocity measures exhibit strong positive correlations with both SFR and stellar mass (Spearman $r \sim 0.7$ -- $0.8$). The tightest correlations, however, are obtained when turbulent and bulk motions are combined into the total kinetic velocity, exceeding those derived from either component individually. In contrast, relations based solely on $v_{\rm out}$ display the largest intrinsic scatter and measurement uncertainties among the three definitions. These results indicate that the full kinetic energy budget of the cool-to-warm gas — rather than the coherent outflow speed alone — more faithfully traces the global star-formation activity of the host galaxy. Since kinetic energy scales as $v^2$, the improved scaling of $v_{\rm tot}$ suggests that feedback couples to the CGM through a combined channel that partitions energy into both coherent outflows and turbulent motions. In this picture, star formation regulates not only the launching of large-scale winds but also the redistribution of energy into the turbulent cascade permeating the multiphase CGM.

We further examine the velocity scaling relations after normalizing each velocity by the halo virial velocity, $V_{\rm vir}$, inferred from the stellar mass–halo mass relation of \citet{Moster2013}, and comparing them with the specific star-formation rate (${\rm sSFR} \equiv {\rm SFR}/M_\star$). After such normalization, the correlations weaken substantially. We find only tentative positive correlations for the first two velocity measures, although their statistical significance has dropped below $3\sigma$. These results suggest that once the depth of the potential well is accounted for, any dependence of cool-to-warm wind kinematics on star formation efficiency is weak, and the dominant imprint of feedback is more closely tied to the cumulative energy injected by feedback over the growth history of the galaxy rather than by the star formation efficiency.

Overall, these results demonstrate that incorporating turbulence into the total kinetic velocity of the gas reveals a tighter and more physically robust connection between CGM kinematics and star formation. While the bulk outflow speed traces the coherent launching of large-scale winds, it neglects the substantial reservoir of kinetic energy stored in random motions within and between clumps. The stronger scaling relations obtained using the total kinetic velocity suggest that the imprint of feedback on the CGM is encoded not only in coherent gas acceleration, but also in the turbulent cascade that redistributes energy and momentum across multiple spatial scales. In this framework, turbulence is not merely a secondary by-product of outflows but a dynamically important channel through which stellar feedback couples to the circumgalactic environment. These findings further underscore the importance of treating turbulence as a fundamental component of galactic wind physics, rather than as a minor perturbation to an otherwise coherent bulk outflow.

\section{The Physical Origin of the Large Turbulent Velocities in Galactic Winds}\label{sec:turb_drive}

In the previous section, we demonstrated that turbulent motions associated with galactic winds play a critical role in shaping both the kinematics and energetics of the circumgalactic gas. These results naturally raise the question of the physical origin of the large turbulent velocities — often reaching $\sim 100$ – $300\,{\rm km\,s^{-1}}$ (see Figure~\ref{fig:vturb_sfr_mstar}) in galaxies with particularly high SFRs — observed in the cool-to-warm wind. In this section, we quantitatively examine several physical mechanisms that may generate and sustain such turbulent motions on circumgalactic scales.

The first possible source of turbulence is the random velocity dispersion arising from gravitational motions within dark matter halos. However, as shown in the lower middle panel of Figure~\ref{fig:vtot_SFR_Mstar}, essentially all galaxies in our sample exhibit $\log\,\langle v_{\rm turb}/V_{\rm vir}\rangle >0$. 
This systematic excess indicates that the observed turbulent velocities substantially exceed the virial velocity scale and therefore cannot be attributed solely to virialized gravitational dynamics. Additional energy injection mechanisms are required to account for the observed magnitude of the turbulence.

In the following subsections, we examine two primary mechanisms that may account for the observed large turbulent velocities in galactic winds. First, stellar feedback (particularly in the form of supernova explosions) can inject substantial mechanical energy into the outflow, driving turbulence as the wind expands and interacts with the ambient medium. Alternatively, turbulence may be generated locally at the interfaces between cool-to-warm clumps and the surrounding hot wind through shear-driven turbulent mixing layers (TMLs). We assess the relative ability of these processes to reproduce the magnitude of the wind turbulent velocities inferred in our sample.

\subsection{Stellar Feedback as the Energy Source of Turbulence}\label{sec:sn_feedback_E}

To evaluate whether stellar feedback can energetically account for the observed turbulent and outflow motions in the galactic wind, we compare the total kinetic energy of the gas, $\dot{E}_{\rm gas}$, with the total energy input from supernovae (SNe). We estimate the mechanical energy injection rate from core-collapse SNe directly from the star formation rate under a standard initial mass function (IMF). Massive stars with $m>8\,{\rm M_\odot}$ explode as core-collapse SNe, yielding a number of SNe per unit stellar mass formed of
\begin{equation}
\eta_{\rm SN}
\;\equiv\;
\frac{\int_{8}^{100}\phi(m)\,dm}
     {\int_{0.1}^{100} m\,\phi(m)\,dm}
\;\approx\;
0.009\;{\rm SN}\;{\rm M_\odot}^{-1}
\end{equation}
for a Kroupa or Chabrier IMF (e.g., \citealt{Madau2014}). The corresponding SN rate is
\begin{equation}
\mathcal{R}_{\rm SN}
=\eta_{\rm SN}\times{\rm SFR}
\quad ({\rm SN\;yr^{-1}})
\end{equation}
Assuming a canonical explosion energy per event of
$E_{\rm SN}\approx10^{51}\,{\rm erg}$, the time-averaged energy injection rate from SNe is
\begin{align}
\dot{E}_{\rm SN}
&= \mathcal{R}_{\rm SN}\,E_{\rm SN}
= \eta_{\rm SN}\,E_{\rm SN}\,{\rm SFR}
\nonumber\\[3pt]
&\simeq 10^{49}\,
\left(\frac{\eta_{\rm SN}}{0.01}\right)
\left(\frac{E_{\rm SN}}{10^{51}\,{\rm erg}}\right)
\left(\frac{{\rm SFR}}{{\rm M_\odot\,yr^{-1}}}\right)
\;{\rm erg\,yr^{-1}}
\label{eq:edotsn_simple}
\end{align}
or equivalently $\dot{E}_{\rm SN}\simeq3\times10^{41}\,{\rm erg\,s^{-1}}$ for ${\rm SFR}=1\,{\rm M_\odot\,yr^{-1}}$.

Not all of this mechanical energy couples efficiently to the outflowing wind. We therefore introduce an effective coupling efficiency $\epsilon$, which encapsulates physical processes such as thermalization, turbulent mixing, and radiative losses. The energy injection rate available to power the CGM wind is then
\begin{equation}
\dot{E}_{\rm SN,\,avail}
=\epsilon\,\dot{E}_{\rm SN}
\label{eq:edotsn_avail}
\end{equation}

\begin{figure}
\centering
\includegraphics[width=0.48\textwidth]{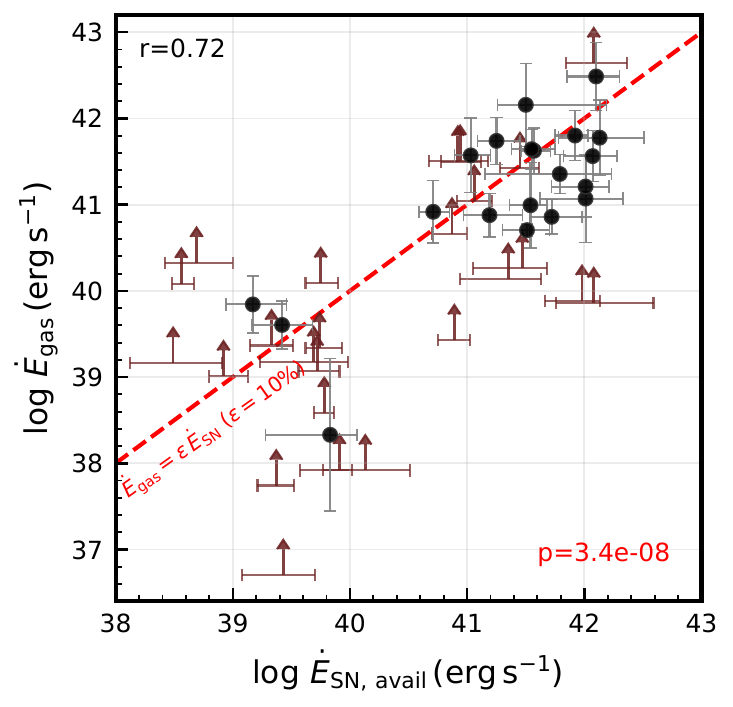}
\caption{
\textbf{Comparison between the total kinetic energy flux of the cool-to-warm wind phase and the supernova energy injection rate inferred from the SFR.} The dashed red line denotes the one-to-one relation $\dot{E}_{\rm gas} = \dot{E}_{\rm SN,\,avail}$ assuming a coupling efficiency of $\epsilon = 10\%$. A strong and statistically significant positive correlation is observed. The majority of galaxies lie around the reference line, indicating that only a modest fraction of the available supernova energy is required to power the observed turbulent and bulk motions of the multiphase wind.}
    \label{fig:Ecool_vs_SFR}
\end{figure}

We compare this quantity to the total kinetic energy flux carried by the gas inferred from our RT modeling, which can be expressed as
\begin{equation}
\dot{E}_{\rm gas}
=\frac{1}{2}\,\dot{M}_{\rm gas}\,
\big(v_{\rm turb}^{2}+v_{\rm out}^{2}\big)
\label{eq:edotcool}
\end{equation}
where $\dot{M}_{\rm gas}$ is the total gas mass outflow rate derived in Section~\ref{sec:mdot_eta}, and $v_{\rm turb}$ and $v_{\rm out}$ represent the turbulent and bulk outflow velocity components of the gas, respectively.

Figure~\ref{fig:Ecool_vs_SFR} compares the inferred total kinetic energy flux of the cool-to-warm wind phase, $\dot{E}_{\rm gas}$, with the supernova energy injection rate available to the wind, $\dot{E}_{\rm SN,\,avail}$. 
The red one-to-one reference line corresponds to the condition
\begin{equation}
\dot{E}_{\rm gas}
=
\dot{E}_{\rm SN,\,avail}
\end{equation}
for an assumed effective coupling efficiency of $\epsilon = 10\%$.

A strong and statistically significant correlation is observed between $\dot{E}_{\rm gas}$ and $\dot{E}_{\rm SN,\,avail}$ ($r=0.72$, $p=3.4\times10^{-8}$), indicating that galaxies with higher supernova energy input systematically exhibit higher kinetic energy fluxes in their winds. The majority of galaxies are distributed around the $\dot{E}_{\rm gas} = 0.1\dot{E}_{\rm SN}$ reference line, implying that only a modest fraction of the available supernova energy needs to couple to the wind in order to account for the observed turbulent and bulk outflowing motions. In other words, supernova feedback appears energetically sufficient to power the kinetic energy carried by the cool-to-warm wind, even for relatively low coupling efficiencies. In reality, most of the supernova energy likely resides in a hot, volume-filling phase that drives the outflow and subsequently entrains and accelerates cooler gas (e.g., \citealt{Veilleux2005, Thompson2024}).

\subsection{Turbulent Mixing as the Energy Source of Turbulence}

Another potential source of energy for the CGM clumps is turbulent mixing
layers (TMLs), which naturally arise at the interfaces between the hot halo gas
and cool-to-warm clouds due to shear-driven instabilities, such as Kelvin--Helmholtz instabilities. In these boundary regions, a fraction of the kinetic energy of the hot phase can be transferred to the cool-to-warm gas through shear work, driving turbulent motions among the clumps. Here we estimate the energy injection rate that can be supplied by TMLs and compare it to the total kinetic energy flux of the CGM gas inferred from our RT modeling.

We model the rate of shear work available to turbulent mixing layers as
\begin{equation}
\dot{E}_{\rm mix}
\simeq \frac{1}{2}\,C_d\,\rho_{\rm hot}\,v_{\rm rel}^{3}\,A_{\rm int}
\label{eq:Emix_power}
\end{equation}
where $\rho_{\rm hot}$ is the mass density of the hot phase, $v_{\rm rel}$ is
the relative shear velocity between the hot gas and the CGM clumps, $C_d$ is a
dimensionless drag coefficient of order $0.1$ -- $1$, and $A_{\rm int}$ denotes
the effective interaction area between the hot and cool phases.

The total interaction area is given by
\begin{align}
A_{\rm int}
&= \zeta\,N_{\rm cl}\,(4\pi R_{\rm cl}^{2}) \nonumber\\[3pt]
&= \frac{4\pi\,\zeta\,F_{\rm V}}{R_{\rm cl}}\,
   \left(R_{\rm out}^{3} - R_{\rm in}^{3}\right)
\label{eq:Aint}
\end{align}
where $\zeta \ge 1$ is a dimensionless factor that parametrizes surface corrugation and unresolved substructure that can enhance the effective interface area.

Only a fraction of the shear work performed within the mixing layers is
ultimately converted into kinetic energy of the cool-to-warm gas in the galactic wind. We therefore define
the effective energy injection rate into the multiphase wind as
\begin{equation}
\dot{E}_{\rm mix,\,avail}
=\epsilon_{\rm mix}\,\dot{E}_{\rm mix}
\label{eq:Emix_avail}
\end{equation}
where $\epsilon_{\rm mix}$ represents the effective energy transport efficiency, accounting for losses due to thermalization, radiative cooling, and incomplete coupling between the hot and cool phases.

We now estimate a representative value of $\dot{E}_{\rm mix,\,avail}$ by adopting typical CGM conditions. Specifically, we assume a hot-phase density of $\rho_{\rm hot}=10^{-28}\,{\rm g\,cm^{-3}}$ (corresponding to $n_{\rm hot}\sim10^{-4}\,{\rm cm^{-3}}$), a drag coefficient of $C_d=0.3$, a relative shear velocity of $v_{\rm rel}=300\,{\rm km\,s^{-1}}$, a smooth interface without surface enhancement $\zeta=1$, a characteristic clump volume filling factor of $F_{\rm V}=0.02$, and an effective mixing efficiency of $\epsilon_{\rm mix}=0.1$. For a wind extending from $R_{\rm in}=1\,{\rm kpc}$ to $R_{\rm out}=10\,{\rm kpc}$ with a characteristic clump radius $R_{\rm cl}=50\,{\rm pc}$, we obtain $\dot{E}_{\rm mix,\,avail} \simeq 2\times10^{39}\,{\rm erg\,s^{-1}}$.

Such a value is significantly lower than the kinetic energy flux inferred for the cool-to-warm phase of the galactic wind in the majority of our sample. Reconciling $\dot{E}_{\rm mix,\,avail}$ with the inferred kinetic energy flux would therefore require substantially larger shear velocities, enhanced effective interface areas, and/or higher hot-phase densities than adopted here. This suggests that, under typical halo conditions, shear-driven turbulent mixing at hot–cool interfaces is unlikely to serve as the primary mechanism sustaining the overall kinetic energy budget of the galactic wind.

Physically, these results imply that although turbulent mixing layers may still play an important role on local scales -- for example, by facilitating mass exchange, metal mixing, and the production of intermediate-temperature gas -- they do not appear capable, on their own, of supplying sufficient energy to maintain the observed level of turbulent and bulk motions in the cool-to-warm wind phase. Instead, the dominant energy input is more plausibly provided by stellar feedback, while mixing layers likely act as a secondary process that redistributes and dissipates energy within the multiphase outflow.

\subsection{Other Possible Mechanisms}\label{sec:other_mec}

While stellar feedback alone appears energetically sufficient to sustain the observed turbulent and outflowing motions of the cool-to-warm phase, additional physical processes may contribute under specific galactic environments or evolutionary stages. For example, cosmic rays (CRs), although ultimately powered by supernova activity, provide a distinct non-thermal transport channel capable of carrying energy and momentum over large distances with relatively weak radiative losses. At present, however, it remains challenging to assess the dynamical importance of CRs using down-the-barrel spectroscopy alone; future spatially resolved IFU observations may provide more direct constraints on their role in shaping CGM wind kinematics.

In addition, large-scale gravitational processes -- such as gas accretion along cosmic filaments, tidal interactions, or galaxy mergers -- can inject turbulence into the CGM through shocks and shear flows. These mechanisms may operate in concert with stellar feedback to influence the structure and energetics of the multiphase halo gas. Nevertheless, further theoretical modeling and spatially resolved observations are required to quantify their relative contributions to the overall kinetic energy budget of galactic winds.

\section{Comparison with Previous Works}\label{sec:prev_works}

To assess the robustness and physical plausibility of the galactic wind properties inferred in this work, we compare our RT modeling results with predictions from hydrodynamic simulations as well as with previous empirical analyses of the CLASSY sample. These comparisons provide independent theoretical and observational benchmarks, enabling a critical evaluation of both the strengths and limitations of our RT modeling framework.

\subsection{Comparison with Hydrodynamic Simulations}\label{sec:compare_hydro}

A wide range of hydrodynamic simulations have been used to investigate galactic winds and their interaction with the surrounding CGM, focusing on observables that closely overlap with those examined in this work. For example, using the FIRE simulations, \citet{Li2021} analyzed two-dimensional column density maps of multiple ions together with gas inflow and outflow rates, providing key insights into the baryon cycle driven by galactic feedback. The FOGGIE simulations have likewise been employed to study the column densities and spatial distributions of ions spanning a broad range of ionization states in galactic halos (e.g., \citealt{Augustin2025}). Similarly, \citet{Piacitelli2025} used ChaNGa simulations to explore ion-specific column density maps and the radial variation of wind-affected halo gas, while \citet{Oren2025} utilized IllustrisTNG simulations to quantify mass-loading factors and the transport of mass and energy through galactic halos.

Collectively, these studies have established an increasingly detailed and consistent picture of the multiphase structure and dynamics of galactic winds and their coupling to the surrounding halo gas. However, comparatively less attention has been devoted to the quantitative characterization of turbulent velocities within the cool-to-warm phases of galactic winds. In particular, directly connecting turbulent motions inferred from observations to the physical definitions of turbulence adopted in hydrodynamic simulations remains non-trivial.
This motivates a detailed comparison between the turbulent velocities inferred from our RT modeling and those measured directly in simulations, which we pursue below.

In recent years, considerable effort has been devoted to quantifying turbulent motions in galactic winds and the CGM, both in full cosmological hydrodynamic simulations and in high-resolution idealized cloud--wind interaction studies. In such analyses, the turbulent velocity dispersion is typically defined as the variance of the three-dimensional velocity field after subtracting coherent bulk motions (e.g., \citealt{Kakoly2025}):

\begin{equation}
\sigma_j^2 = \left\langle \left(u_j - \langle u_j \rangle_\rho \right)^2 \right\rangle
\label{eq:sim_sigma}
\end{equation}
where $j = r, \theta, \phi$, and $\langle u_j \rangle_\rho$ denotes the mass-weighted mean velocity in the corresponding direction. This definition isolates genuinely random motions by explicitly removing coherent bulk flows.

In most hydrodynamic simulations, macroscopic turbulent motions are allowed to develop in all three spatial directions. However, our RT modeling adopts a spherically symmetric wind geometry in which the coherent bulk flow is purely radial. For consistency within this framework, we therefore parameterize the macroscopic random motions as radial perturbations that broaden the radial velocity field. While non-radial velocity components can also broaden line profiles (see Appendix~\ref{app:radial_transverse_degeneracy} for a more detailed discussion), including them introduces additional degeneracies without significantly altering the inferred velocity dispersion. For this reason, we restrict the random velocity component to the radial direction.

In our model, the radial velocity of each clump is decomposed into a coherent bulk outflow component and a stochastic component,

\begin{equation}
u_r = v_{\rm out} + \delta v_r
\end{equation}
where $\delta v_r$ represents macroscopic random motions associated with turbulence. We assume that $\delta v_r$ follows a Gaussian distribution with zero mean and dispersion $\sigma_{\rm cl}$, such that $\langle \delta v_r \rangle = 0$ and $\langle \delta v_r^2 \rangle = \sigma_{\rm cl}^2$. Since the clumps are assumed to be identical in mass, the mass-weighted mean velocity satisfies $\langle u_r \rangle_\rho \simeq v_{\rm out}$. Substituting into Equation~(\ref{eq:sim_sigma}), the radial velocity dispersion becomes

\begin{equation}
\sigma_r^2 = \left\langle \delta v_r^2 \right\rangle_\rho = \sigma_{\rm cl}^2 
\end{equation}
Thus, under our assumption of a Gaussian turbulent velocity field, the variance computed using Equation (\ref{eq:sim_sigma}) is precisely $\sigma_{\rm cl}^2$. Our parameterization thus provides a direct and physically consistent mapping between the turbulent velocity inferred from RT modeling and the velocity dispersion measured in hydrodynamic simulations.

We note, however, although the parameter $\sigma_{\rm cl}$ in our RT framework is mathematically equivalent to the turbulent velocity dispersion defined in hydrodynamic simulations after subtraction of the bulk flow, its physical interpretation remains phenomenological. Real astrophysical turbulence is unlikely to be strictly Gaussian, isotropic, or scale-independent, and is expected to exhibit a cascade across a hierarchy of spatial scales. Our prescription does not attempt to resolve such scale-dependent structure or to model the detailed turbulent cascade. Instead, it provides an effective description of the unresolved macroscopic velocity dispersion that shapes the emergent line profiles. Accordingly, $\sigma_{\rm cl}$ should be interpreted as an effective velocity dispersion that encapsulates the cumulative impact of random bulk motions within the wind, rather than as a literal, scale-resolved model of turbulence.

The large values of $\sigma_{\rm cl}$ inferred in this work are highly consistent with results from hydrodynamic simulations of galactic outflows. For example, using Enzo simulations with kiloparsec-scale spatial resolution, \citet{Schmidt2021} showed that turbulent velocities in the low-redshift CGM can reach $\sim200$ -- $300~\mathrm{km\,s^{-1}}$. At higher resolution, \citet{Schneider2020}, and more recently \citet{Schneider2024}, employed the Cholla Galactic OutfLow Simulations (CGOLS) to model multiphase outflows driven by supernova feedback in disk galaxies, achieving parsec-scale resolution that resolves the internal structure and kinematics of cool outflowing clouds. 
In these simulations, the radial outflow velocity profiles exhibit substantial dispersion at fixed radius, with characteristic values reaching $\sim 200~\mathrm{km\,s^{-1}}$, closely matching the turbulent velocities inferred from our RT modeling. More recently, \citet{Warren2025} used CGOLS to investigate the detailed kinematics of cool clouds embedded in galactic winds, finding large internal cloud velocity dispersions ($\sim10$ -- $10^{2}\,\mathrm{km\,s^{-1}}$) as well as significant cloud--cloud relative velocities ($\gtrsim10^{2}\,\mathrm{km\,s^{-1}}$). 
These values are fully consistent with both the microscopic turbulent broadening parameter ($b_{\rm D,\,cl}$) and the macroscopic clump velocity dispersion ($\sigma_{\rm cl}$) derived from our RT modeling.

Another recent work by \citet{Kakoly2025} used FIRE cosmological zoom-in simulations found that turbulent pressure generally dominates over thermal pressure in the inner CGM of sub-$L^\star$ galaxies, with turbulent velocities reaching $\sim200~\mathrm{km\,s^{-1}}$. 
They further predicted that such strong turbulence would lead to large EWs ($\gtrsim1$\,\AA) in UV absorption lines. While this prediction is not directly comparable to our analysis — since in our RT framework the absorption line strength depends on a combination of factors including clump kinematics, column densities, and covering fractions, rather than scaling simply with $\sigma_{\rm cl}$ — the broader conclusion that turbulence plays a dominant role in shaping CGM kinematics is highly consistent with our findings.

The large turbulent velocities inferred in this work further underscore the need for expanded numerical studies of turbulent mixing layers in the CGM. Previous investigations (e.g., \citealt{Ji2019, Fielding2020, Tan2021, Tan2021b, Bustard2022, Chen2023, Zhao2023, Abruzzo2024, Blackburn2024, Das2024, Hidalgo-Pineda2025, Ghosh2025, Marin-Gilabert2025}) have largely focused on regimes in which the cool gas is transonic or only mildly supersonic. In contrast, our results indicate that in many galaxies in our sample, the inferred macroscopic turbulent velocities can approach or exceed the sonic speed of the cool-to-warm phase. This highlights the importance of extending future studies to explore turbulent mixing and cloud–wind interactions in higher Mach-number regimes (e.g., \citealt{Yang2023}), where the dynamics, energy transfer, and survival of cool gas may differ qualitatively from those in previously studied parameter spaces. 

We conclude this section by examining the velocity structure function (VSF) of our model, which quantifies the scale-dependent variance of the radial velocity field in the galactic wind. The VSF has recently been employed as a useful diagnostic to characterize velocity fluctuations in the CGM in both observational and numerical studies (e.g., \citealt{Chen2024, Chen2025, Hidalgo-Pineda2025, Warren2025}). Here we compute the second-order longitudinal velocity structure function
\begin{equation}
S_2(\ell)
\equiv
\left\langle \left[ v(r+\ell) - v(r) \right]^2 \right\rangle
\end{equation}
where $v(r)$ denotes the radial gas velocity at radius $r$, $\ell$ is the spatial separation between two locations, and the angle brackets indicate an ensemble average over radius within the region of interest. In this work, we use $S_2(\ell)$ as a statistical diagnostic to characterize how velocity fluctuations vary with physical scale in a multiphase, clumpy outflow.

In our model, the radial velocity field at a given point within a clump at radius $r$ is decomposed into three physically distinct components:
\begin{equation}
v(r) = v_{\rm out}(r) + \delta v_{\rm micro}(r) + \delta v_{\rm macro}(r)
\end{equation}
corresponding to the coherent bulk outflow, microscopic turbulent motions internal to individual clumps, and macroscopic random motions among the clumps. Assuming that these components are statistically uncorrelated, the total velocity structure function can be written as the sum of three contributions
\begin{equation}
S_2(\ell)
=
S_{2,\,\rm out}(\ell)
+
S_{2,\,\rm micro}(\ell)
+
S_{2,\,\rm macro}(\ell)
\end{equation}

The outflow term, $S_{2,\,\rm out}(\ell)$, is computed directly from the bulk outflow velocity profile $v_{\rm out}(r)$ given in Equation (2) in Paper I. This term reflects the deterministic velocity gradient imposed by the large-scale wind acceleration and is therefore only important on large spatial scales and becomes negligible at small separations.

The microscopic turbulence term, $S_{2,\,\rm micro}(\ell)$, describes random motions within the clumps and is characterized by the clump Doppler parameter $b_{\rm D,\,cl}$. These motions decorrelate over a characteristic correlation length $L_{c,\,\rm micro}$, which we take to be the mean chord length through a spherical clump,
\begin{equation}
L_{c,\,\rm micro} \simeq \frac{4}{3}\,R_{\rm cl}
\end{equation}
The corresponding contribution to the velocity structure function is modeled as
\begin{equation}
S_{2,\,\rm micro}(\ell)
=
2\,b_{\rm D,\,cl}^2
\left[
1 - \exp\!\left(-\frac{\ell}{L_{c,\,\rm micro}}\right)
\right]
\end{equation}
which increases quadratically at small separations and asymptotically approaches $2\,b_{\rm D,\,cl}^2$ once $\ell \gg L_{c,\,\rm micro}$.

The macroscopic turbulence term, $S_{2,\,\rm macro}(\ell)$, captures random motions among the clumps and is characterized by the clump velocity dispersion $\sigma_{\rm cl}$. Unlike the microscopic turbulent component, the correlation length of the macroscopic motions is expected to vary with radius, reflecting the changing spatial distribution of clumps. Assuming a clump number density profile $n_{\rm cl}(r)\propto r^{-2}$, the characteristic macro-turbulent correlation length is taken to be the mean nearest-neighbor separation in a Poisson distribution
\begin{equation}
L_{c,\,\rm macro}(r)
\propto n_{\rm cl}(r)^{-1/3}
\end{equation}
At a given radius, the macroscopic contribution to the velocity structure function is therefore modeled as
\begin{equation}
S_{2,\,\rm macro}(\ell,r)
=
2\,\sigma_{\rm cl}^2
\left[
1 - \exp\!\left(-\frac{\ell}{L_{c,\,\rm macro}(r)}\right)
\right]
\end{equation}
and the observable $S_{2,\,\rm macro}(\ell)$ is obtained by averaging this expression over radius across the region of interest.

\begin{figure}
\centering
\includegraphics[width=0.45\textwidth]{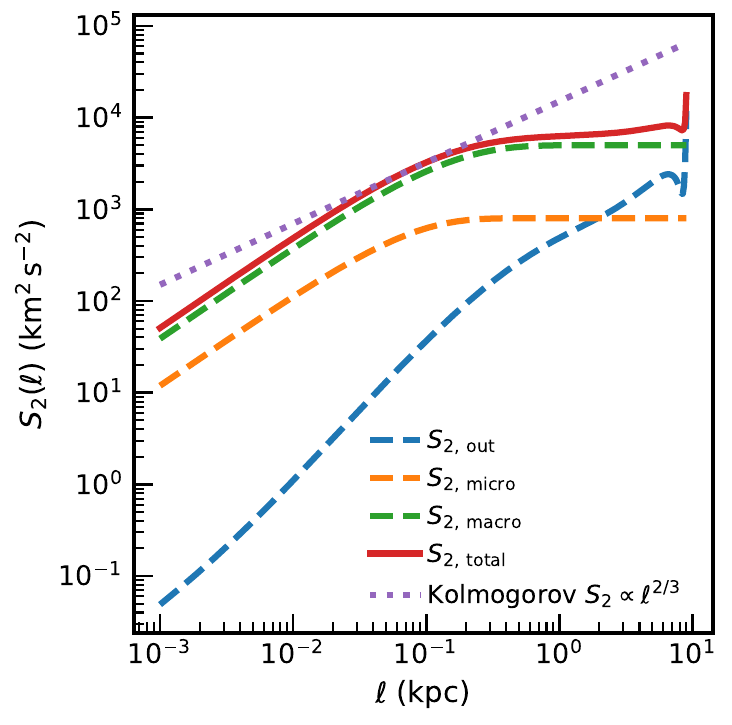}
\caption{\textbf{Second-order longitudinal velocity structure function, $S_2(\ell)$, for the fiducial clumpy CGM model.} The total velocity structure function (solid red curve) is decomposed into contributions from the coherent bulk outflow ($S_{2,\rm out}$; blue dashed), microscopic turbulence ($S_{2,\rm micro}$; orange dashed), and macroscopic random motions
($S_{2,\rm macro}$; green dashed). The model adopts representative parameter values as described in the main text.
Under these conditions, $S_2(\ell)$ is dominated by macroscopic clump motions over most scales
($\ell\sim10^{-2}$--$10~\mathrm{kpc}$). The purple dotted line shows a Kolmogorov scaling, $S_2\propto\ell^{2/3}$, normalized to the total structure function at a characteristic scale.
Although the total $S_2(\ell)$ exhibits an approximately Kolmogorov-like power-law behavior over intermediate scales, this behavior arises from the superposition of finite-correlation-length components in a clumpy CGM wind, rather than from a classical turbulent cascade.
}\label{fig:s2_l}
\end{figure}

Figure~\ref{fig:s2_l} shows the velocity structure function $S_2(\ell)$, decomposed into contributions from different velocity components. We present a fiducial model adopting representative parameters: $F_{\rm V}=0.02$,
$v_0=200~\mathrm{km\,s^{-1}}$, $R=3$, $b_{\rm D,\,cl}=20~\mathrm{km\,s^{-1}}$, $\sigma_{\rm cl}=50~\mathrm{km\,s^{-1}}$, $R_{\rm cl} = 50\,{\rm pc}$, $R_{\rm in} = 1\,{\rm kpc}$, and $R_{\rm out} = 10\,{\rm kpc}$. Under these conditions, $S_2(\ell)$ is dominated by macroscopic clump-clump
motions over nearly the entire range of separations ($\ell\sim10^{-2}$--$10~\mathrm{kpc}$),
while the contribution from microscopic turbulence saturates rapidly at
scales comparable to the clump size and remains subdominant. The bulk outflow term contributes appreciably only at the largest separations
($\ell\gtrsim{\rm kpc}$), where coherent acceleration of the flow becomes
important.

The overall behavior of $S_2(\ell)$ closely resembles that found in recent cloud-wind simulations (e.g., \citealt{Warren2025, Hidalgo-Pineda2025}). In particular, the total structure function exhibits an approximately
Kolmogorov-like power-law behavior over intermediate scales
($\sim10~\mathrm{pc}$ to $\sim1~\mathrm{kpc}$),
with a slope comparable to the Kolmogorov expectation, $S_2\propto\ell^{2/3}$. However, we emphasize that our model does not explicitly incorporate a classical turbulent cascade; instead, the quasi Kolmogorov-like scaling naturally emerges from the superposition of multiple velocity components with several finite correlation lengths, set by the clump size, the clump-clump separation implied by $F_{\rm V}$, and the large-scale outflow geometry. In this sense, the large turbulent velocities and quasi scale-free velocity statistics inferred in galactic winds may not necessarily require a fully developed Kolmogorov cascade. Rather, similar phenomenology may emerge from the superposition of structured, multi-scale bulk motions embedded within an inhomogeneous outflow.

\begin{figure*}
\centering
\includegraphics[width=\textwidth]{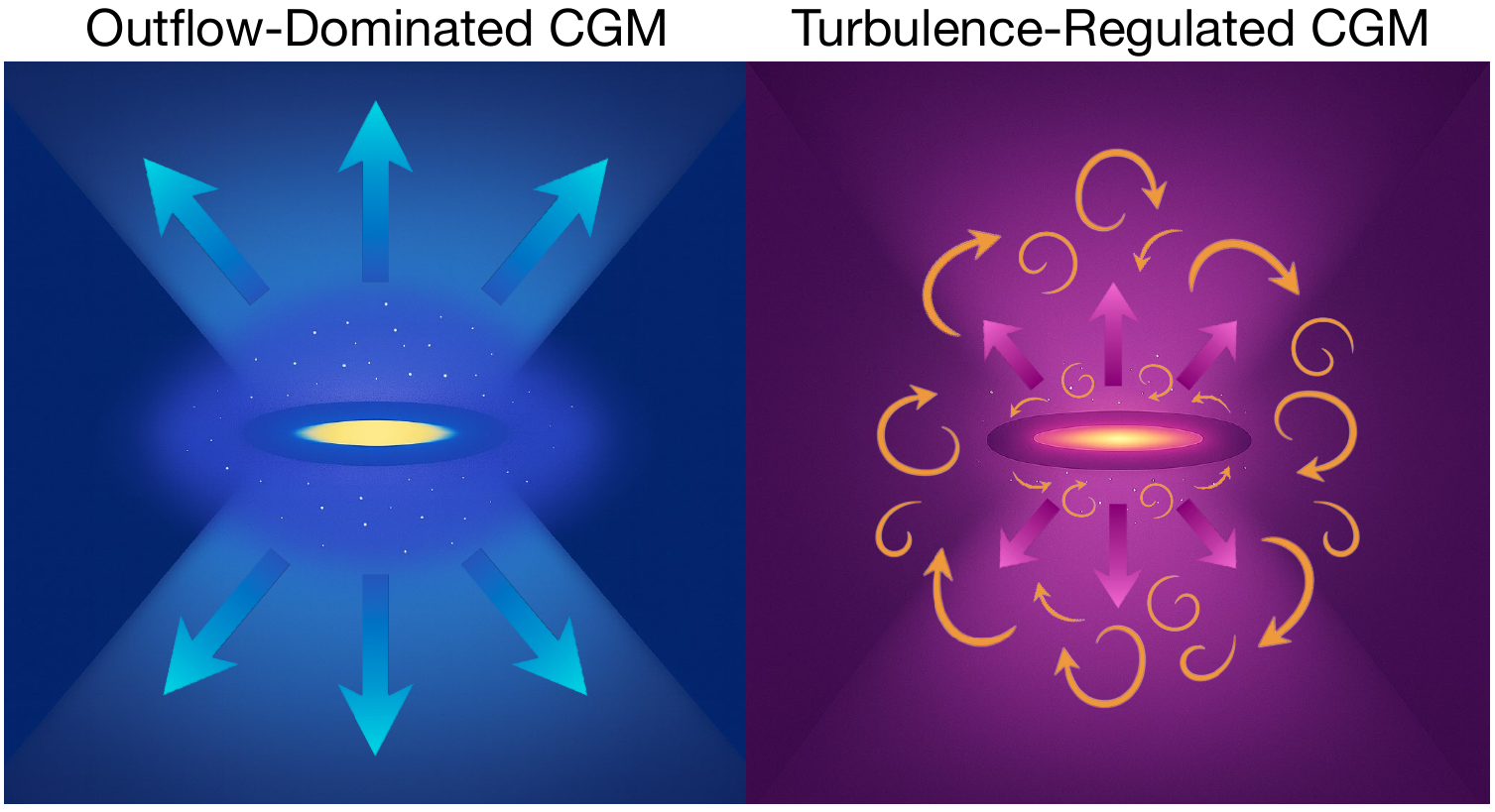}
\caption{\textbf{Schematic comparison between two physical pictures of the CGM.}
\emph{Left: Outflow-dominated CGM.}
In the traditional framework, stellar feedback primarily drives coherent, large-scale bulk outflows that transport mass, momentum, and energy away from the galaxy, while turbulence is typically treated as a secondary by-product of the outflow.
\emph{Right: Turbulence-regulated CGM.}
In the picture emerging from this work, a substantial fraction of feedback energy is instead partitioned into turbulent motions within and among cool-to-warm gas clumps. These stochastic motions play a dynamically significant role in regulating the kinematics, energetics, and structural properties of the CGM, while coherent outflows remain present but no longer necessarily dominate the velocity field. Straight arrows denote coherent bulk outflows, large curved arrows represent macroscopic turbulence, and small curved arrows indicate microscopic turbulence. This schematic highlights the conceptual shift from an outflow-centered interpretation toward a turbulence-regulated view of CGM dynamics.}
\label{fig:turb_CGM}
\end{figure*}

We reiterate that the above decomposition provides an effective statistical description of the velocity field within our RT framework. The parameters $b_{\rm D,\,cl}$ and $\sigma_{\rm cl}$ are inferred from line-profile modeling and do not represent a fully resolved turbulent cascade. The adopted exponential correlation functions and scale-dependent correlation lengths serve as minimal parametric prescriptions for spatial decorrelation, rather than explicit models of scale-dependent turbulence. The resulting velocity structure function should therefore be interpreted as an effective representation of the unresolved velocity field.

At this point, we have demonstrated the critical role of turbulence in the CGM using multiple independent diagnostics, including its kinematics, energetic budget, mass-loading implications, and velocity structure function. Taken together, these results motivate a broader reassessment of the prevailing physical picture of the CGM. Figure~\ref{fig:turb_CGM} presents a schematic comparison between the traditional paradigm and the turbulence-regulated picture emerging from our analysis. In the conventional picture, stellar feedback energy is assumed to primarily drive coherent bulk outflows that transport mass and momentum away from galaxies. By contrast, our results reveal a substantial shift in energy partition, in which a significant fraction of feedback energy is converted into turbulent motions within and among cool-to-warm gas clumps. In this revised framework, turbulence becomes a dynamically significant component that regulates the kinematics, energetics, and structural properties of the CGM. This perspective calls for a re-examination of how feedback energy is distributed and dissipated throughout the galaxy--halo ecosystem.

The results presented here also carry important implications for the design and interpretation of future hydrodynamic simulations of galactic winds and the CGM. In particular, they underscore the need to treat turbulence as an energetically significant component of the multiphase medium. Rather than channeling the majority of feedback energy exclusively into coherent, directed bulk outflows, simulations should consider allocating a substantial fraction of this energy to stochastic velocity perturbations that represent turbulent stirring driven by supernova explosions, radiation pressure, and associated instabilities within the wind.

In addition, our multi-ion analysis provides clear and testable predictions that can be directly evaluated in hydrodynamic simulations. These include examining whether simulations reproduce the relative ordering of column densities and mass outflow rates inferred from different ions (e.g., \SiII, \SiIII, and \SiIV), as well as testing the degree of velocity coherence across multiple ionization states. Our multi-line RT modeling indicates that low- and high-ionization species generally share a common large-scale outflow velocity field, despite tracing gas with distinct thermal and ionization properties. Verifying this multiphase kinematic coherence in hydrodynamic simulations would provide a critical validation of their feedback prescriptions and turbulence modeling.

\subsection{Comparison with Previous Analyses of the CLASSY Sample}\label{sec:compare_prev_work}

\begin{figure*}
\centering
\includegraphics[width=\textwidth]{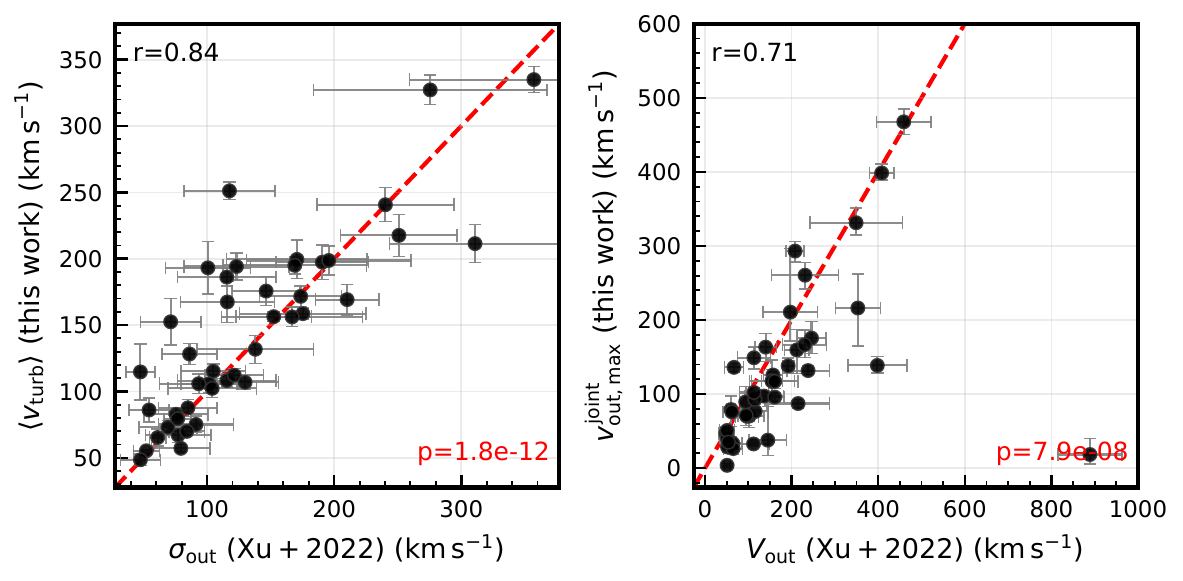}
\caption{
\textbf{Comparison of kinematic parameters inferred from phenomenological double-Gaussian fitting and from full RT modeling.}  \emph{Left panel:} Velocity dispersion of the outflowing Gaussian component inferred by \citet{Xu2022}, converted from FWHM to $\sigma$, plotted against the average turbulent velocity $\langle v_{\rm turb} \rangle$ inferred from our RT modeling. A strong correlation is observed, with most galaxies lying close to the one-to-one relation (dashed line). \emph{Right panel:} Outflow velocity inferred from the centroid of the outflowing Gaussian component in \citet{Xu2022}, compared with the maximum clump outflow velocity $v_{\rm out,\,max}^{\rm joint}$ derived from our joint, multi-ion RT modeling. A significant positive correlation is also present, although in several galaxies the velocities inferred from the double-Gaussian fits exceed those derived from the RT analysis, particularly at the high-velocity end.}
    \label{fig:v_compare_Xu}
\end{figure*}

To place our results in the context of previous analyses of the same galaxy sample, we compare the physical and kinematic properties of the circumgalactic gas inferred from our RT modeling with those derived by \citet{Xu2022}. Although both studies analyze the same CLASSY dataset, they adopt fundamentally different methodological frameworks to infer CGM properties. \citet{Xu2022} combined phenomenological modeling of ultraviolet absorption line profiles with ionization-based analyses of ionic column densities. Specifically, they employed double-Gaussian fitting to characterize the kinematics of the absorbing gas, used a partial covering model (PCM) to infer the column densities of selected ionic species (e.g., \SiII\ and \SiIV), and complemented these measurements with \texttt{CLOUDY} photoionization modeling to constrain additional ions such as \SiIII, \HI, and the total hydrogen column density.

We begin by comparing the kinematic properties inferred by \citet{Xu2022} with those derived from our full RT modeling. In \citet{Xu2022}, the kinematics of UV absorption lines are characterized using a phenomenological double-Gaussian fitting approach, in which the observed profiles are decomposed into two Gaussian components with independent centroids and widths, interpreted as tracing semi-static gas and outflowing material, respectively.

In the left panel of Figure~\ref{fig:v_compare_Xu}, we compare the velocity dispersions derived from the double-Gaussian modeling with the total turbulent velocity inferred from our RT framework. For consistency, the line widths from \citet{Xu2022} are converted from FWHM to velocity dispersion by scaling down by a factor of 2.355. Our quantity $\langle v_{\rm turb}\rangle$ represents the total clump turbulent velocity (including both micro- and macroscopic components), averaged over all available metal ions. We find a tight correlation between the velocity dispersion measured from the double-Gaussian fits and the average turbulent velocity inferred from our RT modeling, with the majority of galaxies lying close to the one-to-one relation. This indicates that the large line widths inferred from phenomenological Gaussian fitting are broadly consistent in magnitude with the turbulent velocities required by physically motivated RT modeling. Our results therefore provide a physical interpretation for the large line widths commonly inferred from UV absorption line profile fitting.

\begin{figure*}
\centering
\includegraphics[width=\textwidth]{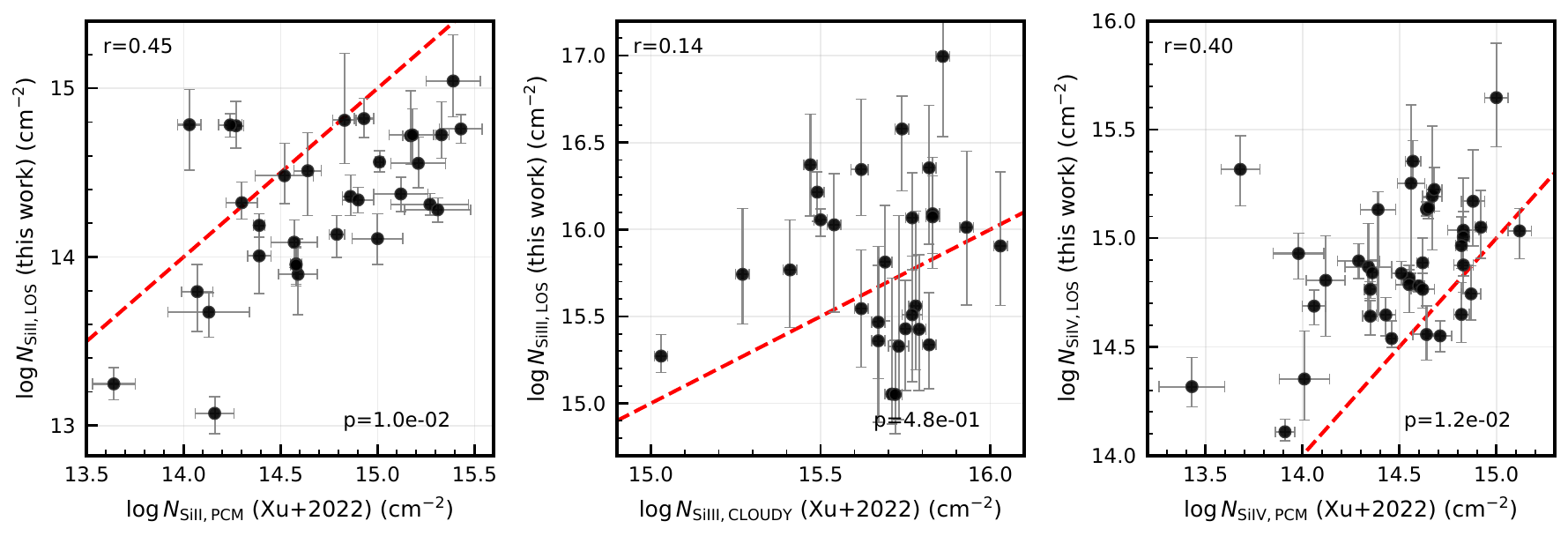}
\caption{\textbf{Comparison of silicon ionic column densities inferred by \citet{Xu2022} and those derived from our RT modeling. }  From left to right, the panels show comparisons for \SiII, \SiIII, and \SiIV, respectively. The dashed red lines indicate a one-to-one relation. \citet{Xu2022} inferred the column densities of \SiII\ and \SiIV\ using a partial covering model (PCM), and subsequently derived \SiIII\ column densities through \texttt{CLOUDY} photoionization modeling.  The \SiII\ column densities inferred from the PCM analysis are systematically higher than those inferred from our RT modeling, whereas the \SiIV\ column densities are systematically lower. The \SiIII\ column densities scatter around the one-to-one relation with no clear systematic offset. These differences reflect the distinct geometric and RT assumptions underlying the two modeling frameworks.}
    \label{fig:Si_compare_Xu}
\end{figure*}

In the right panel of Figure~\ref{fig:v_compare_Xu}, we compare the outflow velocities inferred from the centroid of the blueshifted Gaussian component in \citet{Xu2022} with the maximum clump outflow velocities derived from our joint, multi-ion RT modeling. Here again we observe a significant positive correlation, indicating general agreement between the two methods. However, for several galaxies, the outflow velocities inferred by \citet{Xu2022} are systematically higher than those derived from our RT modeling, particularly at the high-velocity end. These discrepancies likely reflect differences in how characteristic outflow velocities are defined in the two approaches. In our RT modeling, the maximum velocity is typically associated with the location of the absorption trough and is constrained self-consistently across multiple ions. In contrast, when absorption profiles are asymmetric  -- especially when the blue wing rises more gradually than the red wing  -- the centroid of the outflowing Gaussian component must be shifted further blueward in order to reproduce the overall profile shape, leading to systematically larger inferred velocities. Such differences underscore the fundamentally distinct modeling philosophies adopted in this work and in \citet{Xu2022}.

We next compare the ion column densities inferred by \citet{Xu2022} with those derived from our RT modeling. In \citet{Xu2022}, the column densities of low-ionization species such as \SiII\ and \SiIV\ are derived in a velocity-resolved manner using a partial covering model (PCM), in which the optical depth and covering fraction are allowed to vary independently in each velocity bin. The resulting velocity-dependent column densities are then integrated over velocity to obtain total ion column densities. Column densities of additional ions, including \SiIII, \HI, and the total hydrogen column density, are subsequently inferred through \texttt{CLOUDY} photoionization modeling.

As shown in Figure~\ref{fig:Si_compare_Xu}, this comparison reveals several differences between the two approaches. For \SiII, we find only a weak correlation, with the PCM-inferred column densities systematically higher than those derived from our RT modeling by approximately $\sim0.4$ dex. For \SiIV, a similarly weak correlation is present, but in this case the PCM-inferred column densities tend to be lower than the RT-derived values, again by roughly $\sim0.4$ dex. In contrast, the \SiIII\ column densities inferred from \texttt{CLOUDY} photoionization modeling scatter broadly around the one-to-one relation, with no clear systematic offset between the two methods.

These differences in the inferred silicon column densities are not unexpected, given the fundamentally distinct assumptions underlying the two modeling approaches. The PCM approach assumes a velocity-dependent covering fraction and effectively treats absorption components independently in each velocity bin. By contrast, in our multiphase, clumpy RT models, the central source is generally fully covered by an ensemble of clumps, and the emergent line profiles are shaped primarily by RT effects rather than by variations in covering fraction. Systematic offsets in individual ionic column densities therefore arise naturally from these differing physical parameterizations.

\begin{figure*}
\centering
\includegraphics[width=\textwidth]{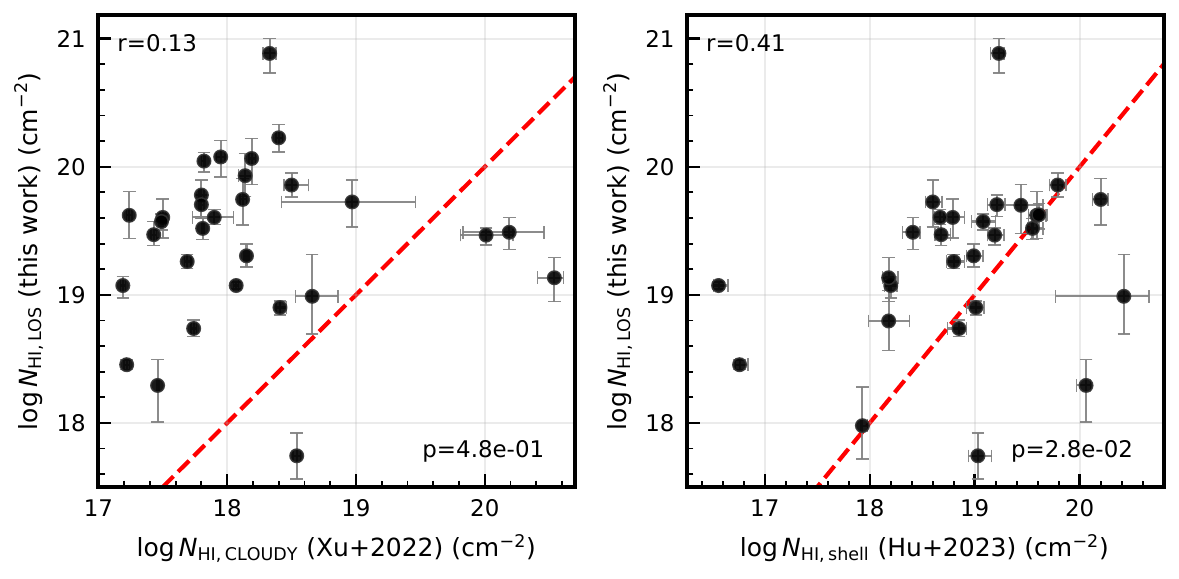}
\caption{
\textbf{Comparison of neutral hydrogen column densities inferred using different modeling approaches.} \emph{Left panel:} Total line-of-sight \HI\ column densities inferred directly from our \lya\ emission RT modeling compared with the \HI\ column densities inferred by \citet{Xu2022}, which are derived indirectly through \texttt{CLOUDY} photoionization modeling constrained by silicon ionic column densities. No statistically significant correlation is observed, and the \citet{Xu2022} values are systematically lower than those derived from our RT analysis. \emph{Right panel:} Comparison between the \HI\ column densities inferred from our clumpy RT modeling and those inferred by \citet{Hu2023}, who modeled \lya\ emission profiles using a monolithic expanding shell geometry. The data points broadly scatter around the one-to-one relation (red dashed line), albeit with substantial scatter. These comparisons highlight the sensitivity of inferred \HI\ column densities to modeling assumptions, and suggest that direct RT modeling of \lya\ emission provides a more internally consistent constraint on the neutral hydrogen content of the wind.}
    \label{fig:NHI_compare_Xu}
\end{figure*}

We next compare the \HI\ column densities inferred by \citet{Xu2022} with the total line-of-sight \HI\ column densities derived directly from our \lya\ emission RT modeling. As shown in the left panel of Figure~\ref{fig:NHI_compare_Xu}, we find no statistically significant correlation between the two measurements, and the \HI\ column densities inferred by \citet{Xu2022} are systematically lower than those obtained from our RT analysis. For comparison, we also include the \HI\ column densities derived by \citet{Hu2023}, who modeled \lya\ emission profiles using a monolithic expanding shell geometry. As shown in the right panel of Figure~\ref{fig:NHI_compare_Xu}, the \HI\ column densities inferred from the shell RT model and from our clumpy RT model exhibit a weak correlation. Although the measurements display considerable scatter, they are distributed roughly symmetrically about the one-to-one relation and do not exhibit a clear systematic offset between the two modeling approaches.

The absence of a correlation with the \citet{Xu2022} values likely reflects the fundamentally different inference pathways adopted in the two studies. In \citet{Xu2022}, the \HI\ column density is derived indirectly: the observed \SiII\ and \SiIV\ column densities are first matched to \texttt{CLOUDY} photoionization models to infer the total hydrogen column density, which is then converted to \HI. Although both \citet{Xu2022} and this work recover the characteristic ordering $N_{\rm SiIII} > N_{\rm SiIV} > N_{\rm SiII}$, which indicates similarly high ionization conditions for silicon, the conversion from total hydrogen to neutral hydrogen in \citet{Xu2022} implicitly assumes a single ionization parameter across all species. In a multiphase wind, however, silicon ions and hydrogen need not trace identical spatial regions or experience the same ionizing radiation field. Differences in ionization potential and shielding can naturally lead to phase-dependent ionization structures. Enforcing a single ionization parameter may therefore introduce systematic biases in the inferred total hydrogen and \HI\ column densities. By contrast, \HI\ column densities inferred directly from RT modeling of \lya\ emission profiles are constrained by the resonant scattering signatures themselves and thus provide a more direct probe of the neutral gas distribution.

\begin{figure*}
\centering
\includegraphics[width=\textwidth]{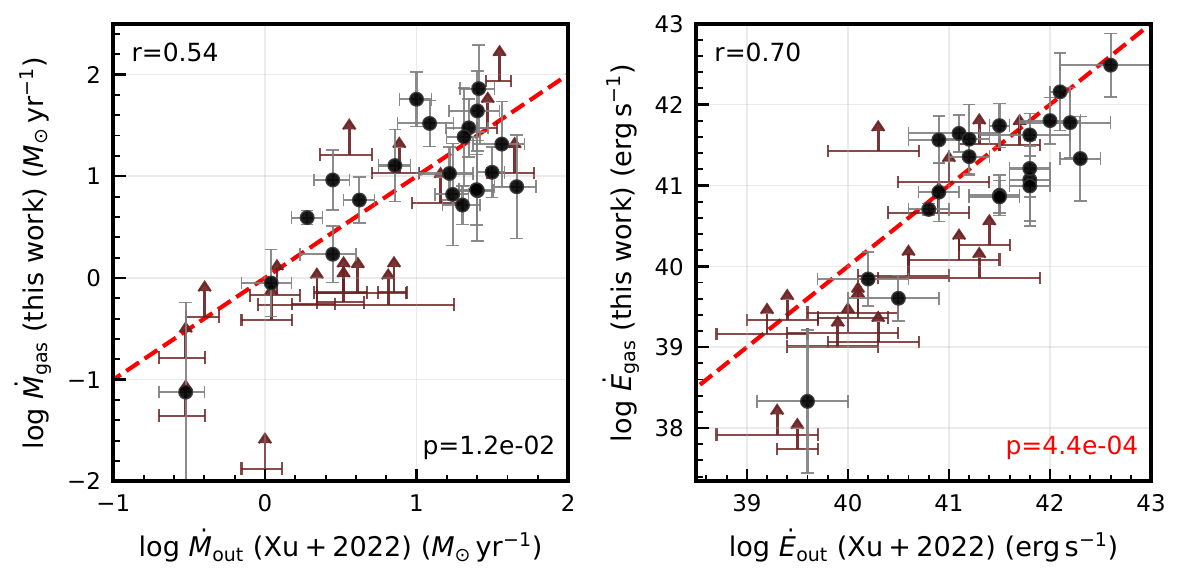}
\caption{
\textbf{Comparison of mass and kinetic energy outflow rates inferred by \citet{Xu2022} and those derived from our RT modeling.}  \emph{Left panel:} For the total gas mass outflow rate, a moderate but statistically significant correlation is observed, with substantial scatter about the one-to-one relation (red dashed line). \emph{Right panel:} For the kinetic energy flux, a stronger and more statistically significant correlation is found, with most systems broadly following the one-to-one relation. 
The larger discrepancies in mass outflow rates likely reflect differences in geometric assumptions and velocity parameterizations between the two analyses. In particular, \citet{Xu2022} compute the mass and energy outflow rates assuming an expanding thin shell with a single, constant bulk velocity. By contrast, our multiphase RT modeling adopts a clumpy geometry with radially varying velocity profiles and explicitly includes both bulk and turbulent velocity components in the kinetic energy budget. The closer agreement in kinetic energy flux therefore suggests that, despite methodological differences, both approaches recover broadly consistent estimates of the global wind energetics when the full kinetic energy content is considered.}
    \label{fig:mdot_edot_compare}
\end{figure*}

Lastly, we compare the mass and kinetic energy outflow rates inferred by \citet{Xu2022} with those derived from our RT modeling in Figure~\ref{fig:mdot_edot_compare}. For the mass outflow rate, we find only a moderate correlation between the two measurements ($r=0.60$, $p=4.9\times10^{-3}$), with substantial scatter about the one-to-one relation. In contrast, the kinetic energy flux exhibits a stronger and more statistically significant correlation ($r=0.76$, $p=1.1\times10^{-4}$). This difference likely reflects the distinct geometric and dynamical assumptions adopted in the two analyses. In \citet{Xu2022}, the mass outflow rate is computed under the assumption of an expanding thin shell with a single, constant outflow velocity and continuous, mass-conserving flow. The corresponding kinetic energy flux is therefore estimated using only the bulk outflow velocity component. By contrast, our multiphase RT modeling assumes a clumpy geometry with radially varying velocity profiles that include both micro- and macroscopic turbulent components. The kinetic energy flux in our framework explicitly accounts for the contribution of turbulent motions in addition to coherent bulk flow. As a result, the inferred energy flux more fully captures the total kinetic energy budget of the wind. The stronger correlation in kinetic energy flux therefore suggests that, despite differences in geometric assumptions and velocity parameterizations, both approaches recover a broadly consistent estimate of the global energetic output of the wind when the full kinetic energy content is considered. In contrast, the mass outflow rate is more sensitive to assumptions regarding geometry and velocity structure, which may naturally lead to larger discrepancies between the two methods.

Taken together, these comparisons demonstrate that, while there is broad consistency in characteristic kinematic amplitudes and ionization trends across different analyses of the CLASSY sample, the detailed physical properties inferred for the circumgalactic gas can exhibit systematic differences depending on the adopted modeling framework. In particular, quantities that depend sensitively on geometric assumptions, velocity parameterizations, or ionization prescriptions -- such as individual ionic column densities, neutral hydrogen content, and mass outflow rates -- can vary significantly between different modeling approaches. Phenomenological and semi-empirical approaches, including double-Gaussian profile decomposition and partial covering analyses coupled with photoionization modeling, are able to reproduce the observed spectra, but may introduce systematic biases in the inferred physical quantities, particularly in a multiphase and kinematically structured wind. In contrast, direct RT modeling of \lya\ and metal line profiles provides a more physically self-consistent means of constraining both the kinematic structure and the neutral hydrogen content of the galactic wind in the CGM. This comparison therefore underscores the importance of incorporating RT effects and realistic gas geometry when interpreting multi-ion UV absorption and emission spectra.

\section{Future Applications}\label{sec:science_appl}

Before concluding, we highlight several major advances enabled by this work that open new avenues for future studies of galactic winds and the CGM. The physically motivated and computationally efficient RT framework, originally developed in Paper I, is applied here to derive new constraints and physical insights. The results presented in this study demonstrate its broad applicability across a wide range of observational and theoretical contexts. Below we outline several promising directions for future development.

\textbf{(1) Extension to higher redshift and JWST observations:} With the advent of \textit{JWST}, rest-frame ultraviolet resonant and fluorescent metal lines (e.g., \CII, \SiII, \MgII, \CIV, \SiIV), as well as \lya, are now routinely detected in galaxies at $z \gtrsim 6$ (e.g., \citealt{Bordoloi2024, Gazagnes2025, Higginson2025, Roberts-Borsani2025, Wu2025}). Our framework can be directly applied to these systems to test whether the turbulence-regulated wind kinematic structure inferred at low redshift also persists into the early Universe, when feedback, gas accretion, and star formation activity are expected to be more extreme. Such applications will provide the first constraints on how the partition of energy between turbulence and coherent outflows evolves across cosmic time.

\textbf{(2) Spatially resolved spectroscopy and 3D mapping:} Recent observations have begun to probe spatially resolved metal emission lines around galaxies (e.g., \citealt{Chisholm2020, Guo2023, Leclercq2024, Kusakabe2024, Vasan2025, Shaban2025}). Ongoing and upcoming integral-field spectroscopic surveys with facilities such as VLT/MUSE, Keck/KCWI, and \textit{JWST}/NIRSpec IFU will deliver increasingly detailed spatially resolved \lya\ and metal-line spectra across individual galaxies and their halos. Our fitting pipeline can be naturally extended to perform spatially resolved RT modeling (see e.g., \citealt{Erb23}), enabling direct mapping of key physical and kinematic parameters across galaxies and their halos. Such analyses will allow direct tests of how galactic wind velocities, turbulent motions, and gas structure vary with radius and environment, offering new insights into the geometry of feedback coupling and the interface regions where turbulent mixing regulates the exchange of mass, momentum, and energy between galaxies and their circumgalactic environments.

\textbf{(3) From down-the-barrel observations to QSO sightlines:} Beyond traditional down-the-barrel spectroscopy of star-forming galaxies, our framework can be naturally extended to absorption line studies of the CGM along background QSO sightlines. Recent work has begun probing metal absorption lines along quasar sightlines at $z > 6$ (e.g., \citealt{Christensen2023, Zou2024, Higginson2025}). By modeling absorption profiles observed at different impact parameters using the same parameterization adopted for the down-the-barrel analysis, this approach establishes a direct connection between the kinematic and physical structures of the galactic wind probed along different sightlines. Such cross-calibration provides a self-consistent framework for interpreting the multiphase wind across a broad range of spatial scales and sightline geometries.

\textbf{(4) Application to large spectroscopic surveys:} The computational efficiency of our fitting pipeline makes it feasible to analyze hundreds to thousands of spectra from both archival and upcoming spectroscopic surveys using instruments such as \textit{HST}/COS, Subaru/PFS, Keck/KCWI, and VLT/MUSE. Current representative programs include the Keck Baryonic Structure Survey (KBSS; \citealt{Chen2020, Prusinski2025}) and the MUSE Quasar-fields Blind Emitters Survey (MUSEQuBES; \citealt{Dutta2025}). Applying this framework to large datasets will enable statistically robust constraints on wind kinematics across diverse galaxy populations, transforming individual-case modeling into population-level inference. 

\textbf{(5) Toward multi-line, multi-phase synergy:} Analyses of rest-frame ultraviolet absorption and emission lines can be naturally combined with optical, infrared, and X-ray diagnostics to construct a unified, multi-phase characterization of galactic winds (e.g., \citealt{Mathur2021, Zhang2024, Veraldi2025, Grayson2025, Xu2025}). In future work, we plan to extend our framework to include recombination lines in the rest-frame optical (such as H$\alpha$ and H$\beta$), as well as additional emission and absorption tracers. Such multi-wavelength synergy will connect the cool, warm, and hot components of the CGM within a single physical framework, providing a comprehensive view of how mass, energy, and metals cycle between galaxies and their surrounding environments.

In summary, the combination of physical self-consistency, computational scalability, and multi-line versatility makes the RT framework presented in this work broadly applicable to both current and next-generation theoretical and observational studies of galactic winds and the CGM. By unifying kinematic modeling and RT across multiple ions and wavelengths, this framework establishes a physically grounded bridge between observations and simulations. It therefore lays the foundation for transforming CGM research from qualitative interpretations toward statistically robust and energetically self-consistent constraints on feedback and the baryon cycle across cosmic time.

\section{Conclusions}\label{sec:conclusion}

In this work, building upon the multi-ion RT modeling framework established in Paper~I, we analyze the inferred kinematic and physical properties of the cool--warm galactic winds in a sample of 50 nearby star-forming galaxies. By systematically examining the kinematics, energetics, and scaling relations derived from the RT modeling outputs, we extract new physical insights into the structure and energy partition of the multiphase CGM wind. Our main conclusions are summarized as follows:

\begin{enumerate}

\item \textbf{Turbulence is dynamically and energetically significant in the cool--warm CGM.} For most galaxies, the inferred macroscopic turbulent velocity is comparable to or exceeds the coherent bulk outflow velocity and often exceed the sound speed of the cool-to-warm phase. The associated macroscopic turbulent pressure frequently dominates over both microscopic pressure and ram pressure. Turbulence therefore constitutes a primary contributor to the kinetic energy and pressure budget of the CGM wind.

\item \textbf{CGM wind kinematics and ion properties scale with host galaxy properties.}  
Outflow velocities, turbulent velocities, ionic column densities, and metal mass outflow rates all increase systematically with stellar mass and star formation rate. These scaling relations demonstrate that stellar feedback plays a central role in shaping the physical state and structure of the galactic wind in the CGM.

\item \textbf{Including turbulence strengthens CGM--galaxy scaling relations.} When turbulent motions are incorporated into the effective velocity and energy budgets, the resulting relations are tighter and more physically interpretable. This behavior indicates that turbulence is not merely a secondary by-product of feedback, but an essential component of the wind dynamics.

\item \textbf{The strengthened scaling relations favor an energy-driven feedback regime.}  
The impact of turbulence on mass loading and kinetic energy flux supports a scenario in which feedback energy, rather than momentum alone, regulates large-scale outflows. Turbulence therefore provides a key channel for feedback energy transport and emerges as a fundamental outcome of stellar feedback that regulates CGM structure, multiphase coupling, and wind energetics.

\item \textbf{Stellar feedback supplies sufficient energy to sustain both turbulence and coherent outflows.}  
The total kinetic energy flux of the cool--warm CGM strongly correlates with the mechanical energy injection rate from star formation. Only a modest coupling efficiency is required to power the observed turbulent motions and bulk outflows. Alternative mechanisms, such as virialized gravitational motions or shear-driven turbulent mixing layers, are energetically subdominant under typical halo conditions.

\item \textbf{Self-consistent RT modeling provides a physically grounded interpretation of phenomenological analyses.}  
While double-Gaussian and partial covering methods broadly reproduce observed line widths, systematic offsets arise in inferred outflow velocities and ionic column densities when compared with self-consistent RT modeling. These differences highlight the sensitivity of derived CGM properties to geometric and ionization assumptions.

\item \textbf{The galactic wind in the CGM exhibits a turbulence-regulated energy partition.}  
Our results indicate that a substantial fraction of stellar feedback energy is stored in stochastic motions within and among cool--warm clumps, rather than being carried exclusively by coherent bulk outflows. In this framework, turbulence emerges as a dynamically important component that shapes the kinematics, energetics, and multiphase structure of the CGM wind.

\end{enumerate}

Overall, our results establish turbulence as a dynamically significant component of the multiphase wind in the CGM and demonstrate that physically grounded, joint RT modeling across multiple ions provides a robust bridge between ultraviolet observations and theoretical models of galactic feedback. By unifying emission and absorption diagnostics over a wide range of ionization states within a single framework, this approach enables a self-consistent interpretation of CGM kinematics and energetics that is not attainable through purely phenomenological analyses or single-line modeling. Leveraging the modeling framework developed in Paper~I, the \texttt{PEACOCK} RT pipeline now offers a systematic pathway toward statistically robust and physically interpretable studies of galactic winds in the CGM, from the local universe to the epoch of reionization.

%% IMPORTANT! The old "\acknowledgment" command has be depreciated. It was
%% not robust enough to handle our new dual anonymous review requirements and
%% thus been replaced with the acknowledgment environment. If you try to 
%% compile with \acknowledgment you will get an error print to the screen
%% and in the compiled pdf.
%% 
%% Also note that the akcnowlodgment environment does not support long amounts of text. If you have a lot of people and institutions to acknowledge, do not use this command. Instead, create a new \section{Acknowledgments}.
\begin{acknowledgments}
We acknowledge the contributions of the CLASSY team members, whose efforts made this project possible. This work was carried out using the Advanced Research Computing at Hopkins (ARCH) core facility (rockfish.jhu.edu), which is supported by the National Science Foundation under grant OAC-1920103. MG acknowledges support from the Max Planck Society through the Max Planck Research Group.
\end{acknowledgments}

%% To help institutions obtain information on the effectiveness of their 
%% telescopes the AAS Journals has created a group of keywords for telescope 
%% facilities.
%
%% Following the acknowledgments section, use the following syntax and the
%% \facility{} or \facilities{} macros to list the keywords of facilities used 
%% in the research for the paper.  Each keyword is check against the master 
%% list during copy editing.  Individual instruments can be provided in 
%% parentheses, after the keyword, but they are not verified.

\vspace{5mm}
\facilities{HST (COS)}

%% Similar to \facility{}, there is the optional \software command to allow 
%% authors a place to specify which programs were used during the creation of 
%% the manuscript. Authors should list each code and include either a
%% citation or url to the code inside ()s when available.

%\software{astropy \citep{2013A&A...558A..33A,2018AJ....156..123A}}

%% Appendix material should be preceded with a single \appendix command.
%% There should be a \section command for each appendix. Mark appendix
%% subsections with the same markup you use in the main body of the paper.

%% Each Appendix (indicated with \section) will be lettered A, B, C, etc.
%% The equation counter will reset when it encounters the \appendix
%% command and will number appendix equations (A1), (A2), etc. The
%% Figure and Table counter will not reset.

\appendix
\section{Radial Scaling of $\dot{M}_{\rm gas}$ and $\dot{E}_{\rm gas}$}
\label{app:radial_scaling}

\begin{figure*}
\centering
\includegraphics[width=0.9\textwidth]{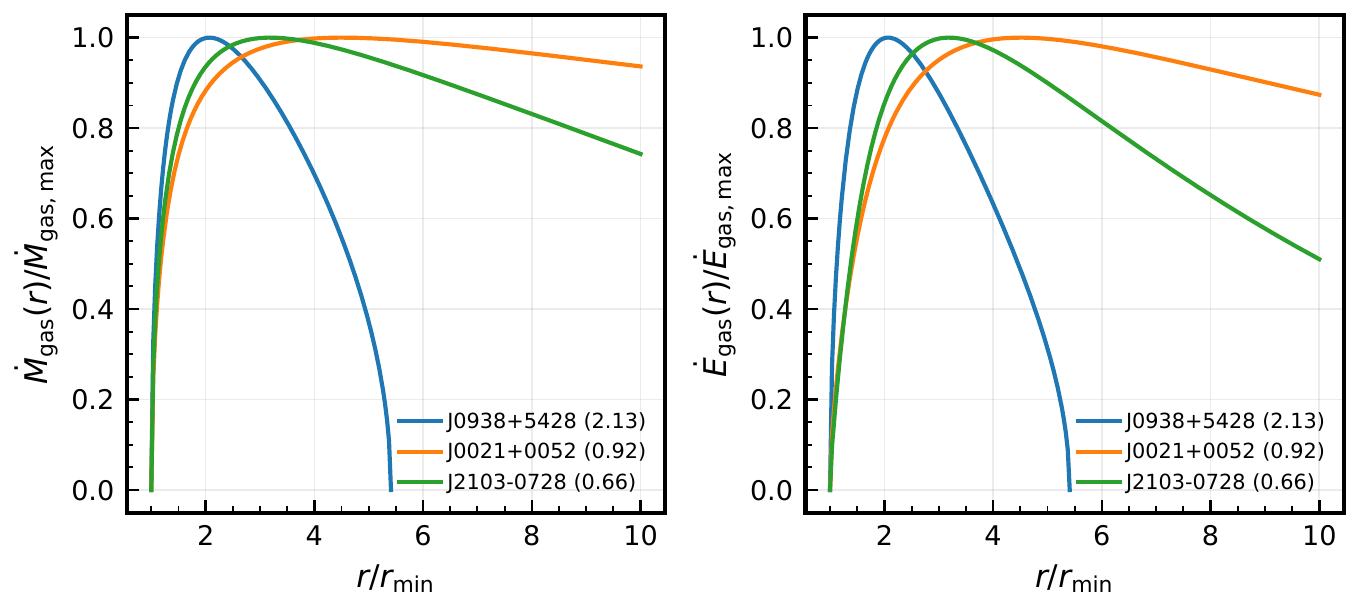}
\caption{\textbf{Normalized radial profiles of the gas mass flux and kinetic energy flux for J0938+5428, J0021+0052, and J2103$-$0728.} \textit{Left:} Normalized mass outflow rate, ${\dot{M}}_{\rm gas}(r)/\dot{M}_{\rm gas,\,max}$. \textit{Right:} Normalized kinetic energy flux, ${\dot{E}}_{\rm gas}(r)/\dot{E}_{\rm gas,\,max}$. 
In each panel, profiles are normalized by their respective maximum values to facilitate comparison of radial shapes. The horizontal axis shows radius in units of $r_{\rm min}$. Numbers in parentheses indicate the ratio $\langle v_{\rm turb} \rangle / v_{\rm out,max}$ for each galaxy.}
\label{fig:app_siii_radial_profiles}
\end{figure*}

In the main text, we adopt the maximum outflow velocity $v_{\rm out,\,max}$ when estimating the total gas mass outflow rate $\dot{M}_{\rm gas}$ and kinetic energy flux $\dot{E}_{\rm gas}$. However, because the outflow velocity $v_{\rm out}(r)$ varies with radius, both $\dot{M}_{\rm gas}$ and $\dot{E}_{\rm gas}$ are in principle radially dependent quantities. In this section, we derive their radial scalings and illustrate representative radial profiles.

From Equations~(\ref{eq:mdot_basic})--(\ref{eq:mdot_gas}), the gas mass outflow rate scales linearly with the radial outflow velocity
\begin{equation}
\dot{M}_{\rm gas}(r) \propto v_{\rm out}(r)
\label{eq:app_mdot_vr}
\end{equation}
so the radial shape of $\dot{M}_{\rm gas}(r)$ directly traces the radial outflow velocity profile. The kinetic energy flux is given by Equation~(\ref{eq:edotcool}):
\begin{equation}
\dot{E}_{\rm gas}(r)
=
\frac{1}{2}
\dot{M}_{\rm gas}(r)
\left[
v_{\rm turb}^2 + v_{\rm out}^2(r)
\right]
\end{equation}
Substituting Equation~(\ref{eq:app_mdot_vr}) yields
\begin{equation}
\dot{E}_{\rm gas}(r)
\propto
v_{\rm out}(r)
\left[
v_{\rm turb}^2 + v_{\rm out}^2(r)
\right]
\label{eq:app_edot_vr}
\end{equation}

Two limiting regimes naturally emerge. When $v_{\rm out}(r) \ll v_{\rm turb}$, the turbulent term dominates and
\[
\dot{E}_{\rm gas}(r) \propto v_{\rm turb}^2\, v_{\rm out}(r)
\]
so the energy flux scales linearly with $v_{\rm out}(r)$, similar to the mass flux. In contrast, when $v_{\rm out}(r) \gg v_{\rm turb}$, the bulk outflow term dominates and
\[
\dot{E}_{\rm gas}(r) \propto v_{\rm out}^3(r)
\]
leading to a much steeper radial dependence.

To compare profile shapes across galaxies with different absolute energetics, we normalize each profile by its maximum value and show the normalized $\dot{M}_{\rm gas}(r)$ and $\dot{E}_{\rm gas}(r)$ in Figure~\ref{fig:app_siii_radial_profiles}. We select three representative systems -- J0938+5428, J0021+0052, and J2103$-$0728 -- with different ratios of $\langle v_{\rm turb} \rangle / v_{\rm out,\,max}$ (2.13, 0.92, and 0.66, respectively).

We find that the mass outflow rate profiles, $\dot{M}_{\rm gas}(r)$, closely track the radial velocity profile $v_{\rm out}(r)$, as expected from the linear scaling in Equation~(\ref{eq:app_mdot_vr}). In contrast, the energy-flux profiles, $\dot{E}_{\rm gas}(r)$, exhibit different behavior depending on the relative importance of turbulent and coherent motions. In the turbulence-dominated case (e.g., J0938+5428), $\dot{E}_{\rm gas}(r)$ approximately follows $v_{\rm out}(r)$, consistent with the linear scaling regime. In systems where the bulk outflow velocity becomes dominant, $\dot{E}_{\rm gas}(r)$ declines more steeply after the initial acceleration phase due to the cubic dependence on $v_{\rm out}(r)$.

These trends demonstrate that while $\dot{M}_{\rm gas}(r)$ primarily reflects the underlying velocity structure, $\dot{E}_{\rm gas}(r)$ captures the dynamical interplay between turbulent and coherent bulk motions. The analytic relations derived above therefore clarify how the relative contributions of turbulence and bulk flow shape the radial energetics of galactic winds.

\section{Degeneracy Between Radial and Transverse Turbulent Components}
\label{app:radial_transverse_degeneracy}

In this work, for simplicity, we adopt a parameterization in which the macroscopic turbulent velocity dispersion is treated as purely radial, i.e.,
\begin{equation}
\boldsymbol{u}
=
u_r \hat{\boldsymbol{r}},
\qquad
\langle u_r^2 \rangle = \sigma_{\rm cl}^2
\end{equation}
with no explicit transverse component. This assumption provides a minimal description of stochastic motions in the clumpy wind and reduces the dimensionality of the parameter space explored in the RT modeling.

Physically, however, it is unlikely that turbulent motions in galactic winds be purely radial. In a realistic multiphase medium, shear instabilities, cloud--cloud interactions, and turbulent mixing layers are expected to generate velocity fluctuations in both radial and transverse directions. A more general velocity decomposition may therefore be written as
\begin{equation}
\boldsymbol{u}
=
u_r \hat{\boldsymbol{r}}
+
u_t \hat{\boldsymbol{t}}
\end{equation}
with dispersions
\begin{equation}
\langle u_r^2 \rangle = \sigma_r^2,
\qquad
\langle u_t^2 \rangle = \sigma_t^2
\end{equation}

However, the frequency redistribution in resonant scattering depends not on this physical-space decomposition itself, but on the projection of the velocity onto the photon propagation direction,
\begin{equation}
u_\parallel = \boldsymbol{u}\cdot \hat{\boldsymbol{k}}_{\rm in},
\qquad
u_\perp = \boldsymbol{u}\cdot \hat{\boldsymbol{e}}_\perp
\end{equation}
which enters the frequency shift per scattering as
\begin{equation}
\Delta x
=
u_\parallel(\mu - 1)
+
u_\perp \sqrt{1-\mu^2}
\end{equation}
where $\mu = \hat{\boldsymbol{k}}_{\rm in}\cdot \hat{\boldsymbol{k}}_{\rm out}$.

The emergent spectrum is determined by the ensemble of escaping photons, and thus effectively constrains the variance of $\Delta x$ over that population. Because the projection from $(u_r,u_t)$ to $(u_\parallel,u_\perp)$ depends on the instantaneous photon direction, and because the angular distribution of escaping photons is set by the numerical RT process, the mapping between $(\sigma_r,\sigma_t)$ and the effective frequency diffusion strength is relatively complex.

\begin{figure}
\centering
\includegraphics[width=0.45\textwidth]{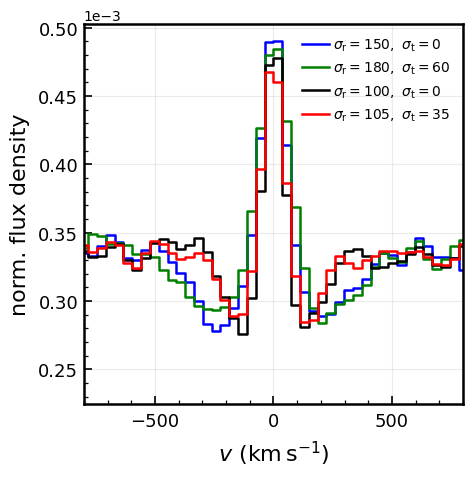}
\caption{\textbf{Comparison of} \SiIII\ \textbf{line profiles obtained with different radial and transverse turbulent velocity decompositions.}The blue and green curves show models with $(\sigma_r,\sigma_t) = (150,0)$ and $(180,60)\,\mathrm{km\,s^{-1}}$, 
respectively, while the black and red curves correspond to $(100,0)$ and $(105,35)\,\mathrm{km\,s^{-1}}$. Despite the different physical-space velocity partitions, each pair produces nearly indistinguishable emergent spectra. This illustrates a degeneracy in which different combinations of radial and transverse turbulent dispersions yield similar projected stochastic velocities along photon propagation directions, and therefore comparable frequency redistribution and escape statistics in the RT. The result demonstrates that the spectra primarily constrain an effective projected turbulent amplitude rather than the detailed three-dimensional anisotropy of the turbulent field.}
\label{fig:siii_sigma_degeneracy}
\end{figure}

This degeneracy has a clear physical manifestation in the absorption line profiles. Increasing the transverse velocity dispersion $\sigma_t$ enhances the frequency diffusion term proportional to $\sqrt{1-\mu^2}$, which tends to redistribute photons away from the frequencies of maximum optical depth and increases the probability that scattered photons are redirected back into the observer’s line of sight. As a result, the absorption trough becomes shallower and the emission peak is reduced due to stronger frequency mixing. To restore the original trough depth, the efficiency with which photons are removed from the emission near the resonance frequency must increase, which can be achieved by increasing the radial velocity dispersion $\sigma_r$. Consequently, we find that an increase in $\sigma_t$ can be compensated by a corresponding increase in $\sigma_r$, yielding nearly identical emergent spectra despite different physical decompositions of the turbulent velocity field.

This degeneracy is illustrated in Figure~\ref{fig:siii_sigma_degeneracy}. The \SiIII\ line profiles obtained with $(\sigma_r,\sigma_t) = (100,0)\,\mathrm{km\,s^{-1}}$
and  $(\sigma_r,\sigma_t) = (105,35)\,\mathrm{km\,s^{-1}}$ are nearly indistinguishable, despite their different physical-space decompositions. A similar degeneracy is also seen for the higher-dispersion pair $(150,0)$ and $(180,60)\,\mathrm{km\,s^{-1}}$. Although these parameter sets differ in their radial and transverse velocity partitions, they generate similar effective projected velocity dispersions within the escaping photon population, leading to nearly identical frequency redistribution statistics and emergent spectra.

This result demonstrates that forward RT modeling primarily constrains an effective stochastic velocity amplitude along the photon propagation direction rather than the detailed three-dimensional anisotropy of the turbulent field. The purely radial turbulence adopted in the main analysis should therefore be interpreted as an effective parameterization of the frequency diffusion strength. Breaking this degeneracy would require observables sensitive to transverse velocity structure, such as spatially resolved spectroscopy or off-axis sightlines.

\section{Best-fit Parameters from RT Modeling}
\label{app:rt_bestfit_tables}

We present two parameter tables derived from the RT modeling in Paper~I: (1) the joint metal-line fitting results and 
(2) the Ly$\alpha$-only fitting results. Tables \ref{tab:joint_metal_fit} and \ref{tab:lya_only_fit} list the best-fit model parameters with their 1-$\sigma$ uncertainties that are adopted throughout the present work. These best-fit model parameters provide the quantitative foundation for the physical diagnostics discussed in this paper. In particular, they serve as the input for deriving key galactic wind properties, including ionic and total gas mass outflow rates, kinematic ratios, and energy-related diagnostics. Parameters such as $F_{\rm V}$, $f_{\rm cl}$, $N_{\rm ion,\,cl}$, $b_{\mathrm{D,\,cl}}$, $\sigma_{\rm cl}$, and the velocity parameters (e.g., $v_0$ and $R$) are directly propagated into the calculations of $\dot{M_{\rm gas}}$, $\dot{E_{\rm gas}}$, and the scaling relations. The parameters listed here therefore constitute the reference model inputs for all subsequent physical inferences in this study.

\setlength{\LTleft}{-10pt}
\setlength{\LTright}{0pt}
\setlength{\tabcolsep}{2pt}
\begin{longtable}{ll@{\hspace{1pt}}cccccccccccc}
\caption{Joint metal-line fitting parameters.}\label{tab:joint_metal_fit}\\
\hline
Galaxy & Ion & $F_{\rm V}$ & $f_{\rm cl}$ & $\log N_{\rm ion,\,cl}$ & $b_{\mathrm{D,\,cl}}$ & $\sigma_{\mathrm{cl}}$ & $r_{\rm pop}$ & $\mathrm{EW}_{\rm int}$ & $b_{\rm max}/R_{\rm out}$ & $\tau_{\rm d,\,cl}$ & $v_0$ & $R$ & $\Delta v$\\
 &  & (\%) &  & (cm$^{-2}$) & (km\,s$^{-1}$) & (km\,s$^{-1}$) &  & (\AA) &  &  & (km\,s$^{-1}$) &  & (km\,s$^{-1}$)\\
\hline
\endfirsthead
\hline
Galaxy & Ion & $F_{\rm V}$ & $f_{\rm cl}$ & $\log N_{\rm ion,\,cl}$ & $b_{\mathrm{D,\,cl}}$ & $\sigma_{\mathrm{cl}}$ & $r_{\rm pop}$ & $\mathrm{EW}_{\rm int}$ & $b_{\rm max}/R_{\rm out}$ & $\tau_{\rm d,\,cl}$ & $v_0$ & $R$ & $\Delta v$\\
 &  & (\%) &  & (cm$^{-2}$) & (km\,s$^{-1}$) & (km\,s$^{-1}$) &  & (\AA) &  &  & (km\,s$^{-1}$) &  & (km\,s$^{-1}$)\\
\hline
\endhead
J0021+0052 & $\mathrm{C\,II}$ & $2.0_{-0.6}^{+0.6}$ & $8.9_{-2.6}^{+2.5}$ & $13.9_{-0.1}^{+0.2}$ & $20_{-8}^{+17}$ & $245_{-36}^{+32}$ & $0.18_{-0.10}^{+0.11}$ & $\cdots$ & $0.75_{-0.20}^{+0.17}$ & $0.04_{-0.02}^{+0.01}$ & $185_{-17}^{+21}$ & $4.5_{-0.4}^{+0.3}$ & $44_{-7}^{+4}$ \\
 & $\mathrm{Si\,II}$ & $1.2_{-0.5}^{+0.7}$ & $5.4_{-2.4}^{+3.2}$ & $13.3_{-0.3}^{+0.3}$ & $11_{-5}^{+10}$ & $264_{-43}^{+26}$ & $0.05_{-0.03}^{+0.04}$ & $\cdots$ & $0.60_{-0.26}^{+0.27}$ & $0.03_{-0.02}^{+0.02}$ & $\cdots$ & $\cdots$ & $\cdots$ \\
 & $\mathrm{Si\,III}$ & $2.3_{-0.7}^{+0.6}$ & $10.5_{-3.1}^{+2.7}$ & $14.5_{-0.3}^{+0.4}$ & $19_{-4}^{+5}$ & $261_{-27}^{+26}$ & $\cdots$ & $\cdots$ & $0.19_{-0.04}^{+0.07}$ & $0.03_{-0.01}^{+0.01}$ & $\cdots$ & $\cdots$ & $\cdots$ \\
 & $\mathrm{C\,IV}$ & $2.1_{-0.7}^{+0.6}$ & $9.5_{-3.1}^{+2.7}$ & $15.1_{-0.4}^{+0.4}$ & $41_{-8}^{+14}$ & $169_{-25}^{+24}$ & $\cdots$ & $\cdots$ & $0.40_{-0.14}^{+0.14}$ & $0.04_{-0.02}^{+0.02}$ & $\cdots$ & $\cdots$ & $\cdots$ \\
 & $\mathrm{Si\,IV}$ & $3.0_{-0.3}^{+0.2}$ & $13.7_{-1.3}^{+0.9}$ & $13.9_{-0.1}^{+0.1}$ & $23_{-4}^{+5}$ & $245_{-18}^{+19}$ & $\cdots$ & $\cdots$ & $0.16_{-0.02}^{+0.02}$ & $0.007_{-0.004}^{+0.007}$ & $\cdots$ & $\cdots$ & $\cdots$ \\
\hline
J0036-3333 & $\mathrm{C\,II}$ & $1.9_{-0.6}^{+0.5}$ & $8.8_{-2.7}^{+2.4}$ & $14.4_{-0.1}^{+0.1}$ & $28_{-7}^{+12}$ & $186_{-14}^{+14}$ & $0.17_{-0.05}^{+0.05}$ & $\cdots$ & $0.18_{-0.03}^{+0.04}$ & $0.06_{-0.03}^{+0.02}$ & $144_{-7}^{+7}$ & $2.8_{-0.1}^{+0.1}$ & $-18_{-3}^{+3}$ \\
 & $\mathrm{Si\,II}$ & $1.5_{-0.3}^{+0.3}$ & $7.0_{-1.3}^{+1.2}$ & $13.3_{-0.1}^{+0.1}$ & $16_{-2}^{+3}$ & $149_{-8}^{+12}$ & $0.03_{-0.01}^{+0.01}$ & $\cdots$ & $0.15_{-0.01}^{+0.02}$ & $0.047_{-0.005}^{+0.007}$ & $\cdots$ & $\cdots$ & $\cdots$ \\
\hline
\end{longtable}

\setlength{\LTleft}{-10pt}
\setlength{\LTright}{0pt}
\setlength{\tabcolsep}{2pt}
\begin{longtable}{ll@{\hspace{1pt}}cccccccccccc}
\caption{Ly$\alpha$ RT modeling best-fit parameters.}\label{tab:lya_only_fit}\\
\hline
Galaxy & Ion & $F_{\rm V}$ & $f_{\rm cl}$ & $\log N_{\rm ion,\,cl}$ & $b_{\mathrm{D,\,cl}}$ & $\sigma_{\mathrm{cl}}$ & $\mathrm{EW}_{\rm int}$ & $b_{\rm max}/R_{\rm out}$ & $\tau_{\rm d,\,cl}$ & $v_0$ & $R$ & $v_{\rm bulk,\,sec}$ & $\Delta v$\\
 &  & (\%) &  & (cm$^{-2}$) & (km\,s$^{-1}$) & (km\,s$^{-1}$) & (\AA) &  &  & (km\,s$^{-1}$) &  & (km\,s$^{-1}$) & (km\,s$^{-1}$)\\
\hline
\endfirsthead
\hline
Galaxy & Ion & $F_{\rm V}$ & $f_{\rm cl}$ & $\log N_{\rm ion,\,cl}$ & $b_{\mathrm{D,\,cl}}$ & $\sigma_{\mathrm{cl}}$ & $\mathrm{EW}_{\rm int}$ & $b_{\rm max}/R_{\rm out}$ & $\tau_{\rm d,\,cl}$ & $v_0$ & $R$ & $v_{\rm bulk,\,sec}$ & $\Delta v$\\
 &  & (\%) &  & (cm$^{-2}$) & (km\,s$^{-1}$) & (km\,s$^{-1}$) & (\AA) &  &  & (km\,s$^{-1}$) &  & (km\,s$^{-1}$) & (km\,s$^{-1}$)\\
\hline
\endhead
J0021+0052 & $\mathrm{H\,I}^{p}$ & $2.9_{-0.9}^{+0.3}$ & $13.3_{-3.9}^{+1.2}$ & $17.9_{-0.1}^{+0.1}$ & $10_{-4}^{+6}$ & $111_{-17}^{+12}$ & $304_{-80}^{+75}$ & $0.07_{-0.01}^{+0.01}$ & $0.01_{-0.01}^{+0.01}$ & $181_{-23}^{+25}$ & $2.8_{-0.2}^{+0.3}$ & $\cdots$ & $39_{-4}^{+4}$ \\
 & $\mathrm{H\,I}^{s}$ & $0.3_{-0.1}^{+0.1}$ & $1.3_{-0.3}^{+0.3}$ & $19.4_{-0.2}^{+0.2}$ & $60_{-26}^{+44}$ & $244_{-29}^{+25}$ & $\cdots$ & $\cdots$ & $0.02_{-0.01}^{+0.01}$ & $\cdots$ & $\cdots$ & $67_{-28}^{+21}$ & $\cdots$ \\
\hline
J0036-3333 & $\mathrm{H\,I}^{p}$ & $2.4_{-0.6}^{+0.4}$ & $10.8_{-2.8}^{+1.6}$ & $18.0_{-0.1}^{+0.1}$ & $11_{-2}^{+6}$ & $32_{-20}^{+14}$ & $105_{-37}^{+47}$ & $0.06_{-0.01}^{+0.02}$ & $0.018_{-0.015}^{+0.001}$ & $190_{-27}^{+21}$ & $2.2_{-0.1}^{+0.2}$ & $\cdots$ & $-4_{-6}^{+14}$ \\
 & $\mathrm{H\,I}^{s}$ & $0.2_{-0.0}^{+0.1}$ & $1.0_{-0.2}^{+0.3}$ & $19.3_{-0.2}^{+0.1}$ & $27_{-16}^{+10}$ & $118_{-17}^{+23}$ & $\cdots$ & $\cdots$ & $0.006_{-0.004}^{+0.010}$ & $\cdots$ & $\cdots$ & $137_{-22}^{+19}$ & $\cdots$ \\
\hline
J0055-0021 & $\mathrm{H\,I}^{p}$ & $2.8_{-0.5}^{+0.3}$ & $12.9_{-2.5}^{+1.5}$ & $19.1_{-0.1}^{+0.1}$ & $65_{-8}^{+9}$ & $62_{-28}^{+37}$ & $17_{-3}^{+5}$ & $0.41_{-0.07}^{+0.11}$ & $0.003_{-0.002}^{+0.003}$ & $79_{-16}^{+17}$ & $3.3_{-0.4}^{+0.7}$ & $\cdots$ & $-41_{-15}^{+16}$ \\
 & $\mathrm{H\,I}^{s}$ & $1.1_{-0.7}^{+0.6}$ & $5.2_{-3.2}^{+2.7}$ & $18.2_{-0.6}^{+0.5}$ & $13_{-6}^{+9}$ & $75_{-48}^{+46}$ & $\cdots$ & $\cdots$ & $0.005_{-0.003}^{+0.004}$ & $\cdots$ & $\cdots$ & $103_{-35}^{+33}$ & $\cdots$ \\
\hline
J0150+1308 & $\mathrm{H\,I}^{p}$ & $2.5_{-0.6}^{+0.6}$ & $11.2_{-2.8}^{+2.7}$ & $18.7_{-0.1}^{+0.1}$ & $55_{-16}^{+11}$ & $23_{-17}^{+29}$ & $27_{-8}^{+9}$ & $0.32_{-0.08}^{+0.09}$ & $0.006_{-0.003}^{+0.003}$ & $105_{-22}^{+29}$ & $3.5_{-0.5}^{+0.6}$ & $\cdots$ & $-58_{-23}^{+56}$ \\
\end{longtable}

\FloatBarrier
\bibliography{sample701}{}
\bibliographystyle{aasjournalv7}
\end{document}